\documentclass{article}

\usepackage{arxiv}

\usepackage[utf8]{inputenc} 
\usepackage[T1]{fontenc}    
\usepackage{hyperref}       
\usepackage{url}            
\usepackage{booktabs}       
\usepackage{amsfonts}       
\usepackage{nicefrac}       
\usepackage{microtype}      
\usepackage{lipsum}		
\usepackage{graphicx}
\usepackage{natbib}
\usepackage{doi}

\title{Temperature and composition disturbances in the southern auroral region of Jupiter revealed by JWST/MIRI}

\author{Pablo Rodriguez-Ovalle*, Thierry Fouchet, Sandrine Guerlet, Thibault Cavalié, Vincent Hue \\
        \textbf{, Manuel López-Puertas, Emmanuel Lellouch, James A. Sinclair, Imke de Pater, }\\
        \textbf{Leigh N. Fletcher, Michael H. Wong, Jake Harkett, Glenn S. Orton, Ricardo Hueso, }\\
        \textbf{Agustín Sánchez-Lavega, Tom S. Stallard, Dominique Bockelee-Morvan, Oliver King, }\\
        \textbf{Michael T. Roman and Henrik Melin}\\
        ...\\
        * * LESIA, Observatoire de Paris\\
        Université PSL, CNRS, Sorbonne Université, Université Paris-Cité\\
        Meudon, France\\
}

\renewcommand{\shorttitle}{\textit{arXiv} Template}

\hypersetup{
pdftitle={A template for the arxiv style},
pdfsubject={q-bio.NC, q-bio.QM},
pdfauthor={David S.~Hippocampus, Elias D.~Striatum},
pdfkeywords={First keyword, Second keyword, More},
}

\begin{document}
\maketitle


\textbf{keypoints}
\begin{itemize}
    \item The homopause is spatially variable within the polar region and highest within the auroral oval.
    \item The atmosphere inside the Southern Auroral Oval at 1 and 0.01 mbar shows a warming compared with non-auroral regions.
    \item The C$_2$H$_2$ abundance is enhanced inside the Southern Auroral Oval at 0.1 and 7 mbar, and C$_2$H$_6$ shows an increase polewards.
\end{itemize}

\,

\textbf{Key Words:}
\begin{itemize}
    \item Planetary atmospheres
    \item Spectroscopy
    \item Infrared astronomy
    \item Planetary polar regions
\end{itemize}

\begin{abstract}
[Jupiter’s south polar region was observed by JWST/Mid-Infrared Instrument in December 2022. We used the Medium Resolution Spectrometer mode to provide new information about Jupiter’s South Polar stratosphere. The southern auroral region was visible and influenced the atmosphere in several ways: i) In the interior of the southern auroral oval, we retrieved peak temperatures at two distinct pressure levels near 0.01 and 1 mbar, with warmer temperatures with respect to non-auroral regions of 12$\pm$2 K and 37$\pm$4 K respectively. A cold polar vortex is centered at 65$^\circ$S at 10 mbar. ii) We found that the homopause is elevated to 590$^{+25}_{-118}$ km above the 1-bar pressure level inside the auroral oval compared to 460$^{+60}_{-50}$ km at neighboring latitudes and with an upper altitude of 350 km in regions not affected by auroral precipitation. iii) The retrieved abundance of C$_2$H$_2$ shows an increase within the auroral oval, and it exhibits high abundances throughout the polar region. The retrieved abundance of C$_2$H$_6$ increases towards the pole, without being localized in the auroral oval, in contrast with previous analysis \cite{SinclairII}. We determined that the warming at 0.01 mbar and the elevated homopause might be caused by the flux of charged particles depositing their energy in the South Polar Region. The 1-mbar hotspot may arise from adiabatic heating resulting from auroral-driven downwelling. The cold region at 10 mbar may be caused by radiative cooling by stratospheric aerosols. The differences in spatial distribution seem to indicate that the hydrocarbons analyzed are affected differently by auroral precipitation.]
\end{abstract}

\section*{Plain Language Summary}
[JWST/Mid-Infrared instrument observed Jupiter’s south polar region in December 2022. The instrument acquired spectroscopic data in the mid-infrared part of the spectrum, which is sensitive to the temperature of the atmosphere and the chemical abundances. These observations revealed that within the auroral oval there are two regions of high temperatures located at two different altitudes. These are presumably caused by two different phenomena: direct heating from the incoming charged particles in the aurora at a height of 0.01 mbar and adiabatic heating in downdrafts at lower levels. A decrease in temperature was also observed as we approached the South Pole, probably caused by a cold polar vortex associated with stratospheric hazes. We found that the altitude of the homopause (the limit between the well-mixed part of the atmosphere and the part where molecules are separated according their specific weight) is altered by the auroras, being up to 100 km higher in the auroral region. The atmospheric abundances of acetylene and ethane showed an enrichment of acetylene within the auroral oval, and of ethane at the pole, which may indicate that these molecules are not affected in the same way by the energy input of the aurora.]

\section{Introduction}
\label{sec_introduction}

The polar regions of Jupiter's atmosphere are affected by electron and ion precipitation and Joule heating originated by the Jovian magnetosphere. This energy deposition in the atmosphere causes an increase in spectral emission in X-rays \cite{Gerard2014}, UV \cite{Greathouse2021}, near-infrared \cite{Castagnoli2022Jul} and mid-infrared \cite{Caldwell1980Dec,Livengood1993}, and in the millimeter range \cite{Cavalie2023Sep}. In addition to the observable aurora, the consequences of these precipitations have a considerable impact on the thermal structure and the chemical composition of the giant planet’s atmosphere.

One of the main effects of charged particle precipitation is atmospheric warming through Joule heating. \cite{SinclairI} and \cite{SinclairII} studied the thermal structure in the North Polar Region (NPR), using Cassini-CIRS and Voyager-IRIS dataset, respectively, and ground-based observations from IRTF-TEXES. Both studies consistently identified two hotspots in the NPR, located at two different pressure levels inside the auroral oval. The first hotspot was located at 0.01 mbar and is attributed to a downward extension of the thermosphere, heated by Joule heating caused by the particle precipitation itself. The second hotspot, located at 1 mbar, is more puzzling. \cite{SinclairI} favored two possible explanations: adiabatic heating caused by a local downwelling driven by charged particle precipitation or a radiative heating driven by aerosols produced by auroral precipitation, though they have since ruled out the latter \cite{Sinclair2023Dec}. Stratospheric haze layers have indeed been inferred at high latitudes from ground-based Near Infrared (NIR) spectra and Cassini ISS images \cite{Zhang2013Sep} and could significantly warm the stratosphere as suggested by \cite{Zhang2015Dec}, although the peak of aerosol density in the southern hemisphere was measured around 10 -- 20 mbar, deeper than the hotspot located at 1~mbar.
 
The auroral heating is also subjected to temporal variations. Recent studies by \cite{Sinclair2023Apr} attributed the temperature variability in data obtained with the TEXES instrument on the Gemini 8.1-m telescope to magnetospheric compression caused by varying solar wind activity. They found that during a compression event in the magnetosphere, the dusk side of the northern auroral oval at 0.01 mbar warms up, whereas temperatures at 1 mbar in the same horizontal location remain practically unchanged. In the same work, the South Polar Region (SPR) was also observed, and the respective auroral effects in the NPR and SPR were compared. They inferred a difference in the thermal profile between the NPR and the SPR, with the 1-mbar hotspot being more vertically extended down to higher pressure levels (from 1 to 10 mbar) in the SPR. In the NPR, the extent of the lower-altitude hotspot ranges from 1 to 4.7 mbar \cite{Flasar2004,Fletcher2016,SinclairI, Odonoghue2021}.

A second effect of auroral precipitation is the possible variation of the homopause pressure level reported by previous studies. \cite{Parkinson2006}, using Cassini UVIS observations, and \cite{Sinclairhomopause} using IRTF-TEXES observations of the H$_2$ S(1), CH$_3$ and CH$_4$ emission features at 587, 606 and 1248 cm$^{-1}$, found a homopause located at higher altitudes within the auroral region than in neighboring regions. In contrast, \cite{Kim2017Jan}, by combining ground-based observations of CH$_4$ $\nu_3$ and $\nu_4$ lines, found a homopause localized at higher altitudes at lower latitudes than in auroral regions. To obtain this result, \cite{Kim2017Jan} used a non-Local Thermodynamic Equilibrium (non-LTE) CH$_4$ radiative transfer code that has recently been revisited by \cite{Sanchez-Lopez2022Jun}, who used ISO/SWS observation of the fundamental and hot $\nu_3$ CH$_4$ band and inferred a methane homopause pressure level in the equatorial region compatible with that inferred in non-auroral polar regions by \cite{Sinclairhomopause}. Therefore, the question regarding a possible displacement of the homopause in the polar regions of Jupiter remains an ongoing debate. Regarding the mechanism behind the possible upward shift of the homopause, \cite{Sinclairhomopause} pointed to an increased mixing generated by auroral driven heating at higher altitudes that would transport hydrocarbons to higher altitudes, thus changing the level of the homopause compared to nearby non-auroral regions.

Auroral precipitation is also thought to alter hydrocarbon chemistry. Specifically for C$_2$ hydrocarbons, using Cassini-CIRS observations at a planetary scale, \cite{Nixon2010Nov}  measured a meridional profile of C$_2$H$_6$ abundance that slightly increased towards the Polar Regions between 1 and 10 mbar, and a C$_2$H$_2$ profile significantly decreasing polewards between 7 and 0.1 mbar. \cite{Fletcher2016} and \cite{Melin2018} reached similar conclusions using ground-based IRTF-TEXES spectra. On a more local scale, an increase in thermal infrared emissions has been observed in several spectral bands belonging to C$_2$H$_2$, C$_2$H$_4$, and C$_2$H$_6$ within the auroral region \cite{Kostiuk1989May,Drossart1993,Sinclair2023Apr}. Although the enhancement of radiance towards the poles can be correlated with temperature enhancements, these studies suggest that charged particle precipitation may also influence the abundance of these species within the auroral region.

Using their retrieved thermal structure, \cite{SinclairI,SinclairII,Sinclair2023Apr} obtained the abundance of C$_2$H$_2$ and C$_2$H$_6$ hydrocarbons within and outside the auroral oval from space-borne and ground-based dataset. The first two studies mostly addressed the NPR, with the Southern Auroral Oval being hardly sampled, while the third study addressed both the North and South Polar Regions. For the NPR, C$_2$H$_2$ was more abundant within the auroral oval in the three studies. However, the specific pressure level at which this increase in abundance occurred varied from one study to another. While \cite{SinclairI,SinclairII} retrieved an abundance increase at pressures ranging from 0.01 to 4 mbar, \cite{Sinclair2023Apr} found a noticeable increase only between 1 and 4 mbar. For C$_2$H$_6$, the measurements obtained in the three studies are different. While \cite{SinclairI} measured a clear increase at 4.7 mbar, \cite{SinclairII} reported a depletion. The most recent study by \cite{Sinclair2023Apr} did not indicate a clear variation in this regard.
For the SPR, \cite{Sinclair2023Apr} found that both acetylene and ethane showed an increase in their abundance within the Southern Auroral Oval compared to the latitudes near the equator, but at different pressure levels. The enhancement was larger around 1 mbar for C$_2$H$_2$, and around 5 mbar for C$_2$H$_6$. Nevertheless, they cautioned that observing the Southern Oval was difficult even with a large telescope and that their analysis may be affected by insufficient spatial resolution. 

To explain the different behavior of ethane and acetylene in the Polar Regions, \cite{SinclairI,SinclairII} used the chemical model proposed by \cite{DeLaHaye2008a}. This model suggests that ion-neutral chemistry preferentially enhances the production of unsaturated hydrocarbons over that of saturated hydrocarbons. This enhancement of unsaturated hydrocarbon is proposed to be diffused downward within the auroral oval, leading to a local C$_2$H$_2$ maximum. Outside the oval, neutral photochemistry converts C$_2$H$_2$ into C$_2$H$_6$. However, a previous ion-neutral chemical model was proposed by \cite{Kim1994}, in which C$_2$H$_2$ is preferentially destroyed by ion-neutral chemical reactions. This possibility was invoked by \cite{Hue2018Jun} to explain the decoupling in the ethane and acetylene equator-to-pole meridional distributions reported by \cite{Nixon2010Nov}. Indeed, using a 2D transport-chemical model, \cite{Hue2018Jun} showed that neither photochemistry nor a combination of diffusive and advective transport could reproduce the anti-correlated ethane and acetylene meridional distributions seen in the Cassini-CIRS data. Given constraints brought by the temporal monitoring of post SL9-species \cite{Moreno2003Aug,Griffith2004Jul,Lellouch2006Oct}, they concluded that an additional C$_2$H$_2$ loss mechanism was required to explain both C$_2$H$_2$ and C$_2$H$_6$ meridional trends. Alternatively, if no meridional diffusion and transport processes are included in the 2D model, then an additional C$_2$H$_6$ production process is required. One should remember that \cite{Hue2018Jun} interpreted retrievals from the Cassini-CIRS observations, which purposely excluded the longitudinal range of the auroral region at high latitudes \cite{Nixon2010Nov}.

The situation is further complicated because diffusive transport and advection can redistribute species on timescales shorter than their typical chemical lifetime \cite{Lellouch2006Oct}. In this context, spectroscopic observations, simultaneously covering several hydrocarbon species, and sampling the polar regions at a high spatial scale, are essential to better constrain the role of ion-neutral chemistry and transport in auroral regions.

In this study, we present an analysis of the thermal and chemical structure of the Jovian stratosphere in the SPR of Jupiter. We use observations from the James Webb Space Telescope (JWST), specifically from the Mid InfraRed Instrument - Medium Resolution Spectroscopy (MIRI-MRS) \cite{MIRI,MIRIfocal}, obtained on December 24th 2022. MIRI-MRS allows us to combine high spatial resolution and mid-spectral resolution, as well as a simultaneous coverage of the full 555 -- 2080 $\mu$m wavelength range. MIRI spectral resolving power of $\sim3700$ for channel 2 allows us to measure the temperature from 20 mbar to 0.01 mbar, the C$_2$H$_2$ abundance between 3 and 0.1~mbar, and the C$_2$H$_6$ abundance from 5 mbar to 1 mbar, inside and outside the Southern Auroral Oval. It also allowed us to infer the homopause level inside and outside the oval.

We have structured the article as follows: in Section \ref{sec_observations}, we provide a comprehensive presentation of the challenges associated with utilizing the MIRI-MRS instrument with bright objects. We also describe in detail the data reduction process. In Section \ref{sec_data}, we explain the methodology used and the procedure necessary to retrieve the homopause pressure level, the temperature, and the abundances of chemical species, as well as the error analysis of the measurements. Section \ref{sec_results} presents the results on the location of homopause, the 3D thermal structure, and the 3D hydrocarbon distribution. In Section \ref{sec_discussion}, we discuss and compare our results with previous studies, a summary of our is presented in Section \ref{sec_conclusion}.

\section{Observations}
\label{sec_observations}
The MIRI instrument onboard JWST \cite{Rieke2015Jul} observed Jupiter's SPR on 2022 December 24, as part of the $\# 1373$ Early Release Science program (PI: Imke de Pater and Thierry Fouchet). The MRS mode was adopted for these observations. This mode uses 4 integral field units (IFU) that can observe the planet simultaneously, covering between them the spectral range of 347.2 -- 2080 cm$^{-1}$. Each IFU probes a specific spectral range and has an angular resolution and Field of View (FOV) tailored to its specific diffraction limit. These four IFUs are coaligned and cover the spectral ranges from 2080 -- 1307.2, 1331.5 -- 854, 865.8 -- 554.9, and 564.6 -- 347.2 cm$^{-1}$, respectively \cite{MIRI}. Furthermore, each of these 4 ranges (from now on referred to as Channels 1 -- 4) are further divided into three sub-bands ('SHORT', 'MEDIUM' and 'LONG', see Table \ref{table1}). A given sub-band is acquired simultaneously by the four IFUs. Therefore, three successive exposures are required to sample the full spectral range. As a result, our observations yielded 12 different hyperspectral cubes with different temporal, spectral, and spatial sampling, which needed to be projected onto Jupiter's disk. A more in-depth explanation of the spatial registration is detailed in Section \ref{navigation}. Table \ref{table1} summarizes the main information of the dataset used for our analysis.

\begin{table}
\tiny
\caption{JWST MIRI-MRS Observations of Jupiter's South Polar Region.}
\resizebox{\textwidth}{!}{%
\begin{tabular}{cccccc}
\hline
\begin{tabular}[c]{@{}c@{}}Central \\ longitude\\ (SIII W)\end{tabular} & \begin{tabular}[c]{@{}c@{}}Time (UT)\\ on 2022 Dec 24\end{tabular} & Filter & \begin{tabular}[c]{@{}c@{}}Wavenumber\\ range (cm$^{-1}$)\end{tabular} & \begin{tabular} [c]{@{}c@{}}Max. Exposure\\ time (s)\end{tabular} & Dithers \\ \hline
343.5 & 13:26:50 & CH1-SHORT & 2040-1742 & 527.258 & 2 \\
 &  & CH2-SHORT & 1331-1140 & 527.258 & 2 \\
 &  & CH3-SHORT & 866-742 & 316.355 & 2 \\
 &  & CH4-SHORT & 565-477 & 105.452 & 2 \\ \hline
332.5 & 13:09:59 & CH1-MEDIUM & 1766-1508 & 527.258 & 2 \\
 &  & CH2-MEDIUM & 1153-987 & 527.258 & 2 \\
 &  & CH3-MEDIUM & 750-642 & 316.355 & 2 \\
 &  & CH4-MEDIUM & 483-408 & 105.452 & 2 \\ \hline
323.5 & 12:53:18 & CH1-LONG & 1531-1307 & 527.258 & 2 \\
 &  & CH2-LONG & 1000-855 & 527.258 & 2 \\
 &  & CH3-LONG & 649-556 & 316.355 & 2 \\
 &  & CH4-LONG & 413-358 & 105.452 & 2 \\ \hline
71 & 15:49:42 & CH1-SHORT & 2040-1742 & 527.258 & 2 \\
 &  & CH2-SHORT & 1331-1140 & 527.258 & 2 \\
 &  & CH3-SHORT & 866-742 & 316.355 & 2 \\
 &  & CH4-SHORT & 565-477 & 105.452 & 2 \\ \hline
60 & 15:33:03 & CH1-MEDIUM & 1766-1508 & 527.258 & 2 \\
 &  & CH2-MEDIUM & 1153-987 & 527.258 & 2 \\
 &  & CH3-MEDIUM & 750-642 & 316.355 & 2 \\
 &  & CH4-MEDIUM & 483-408 & 105.452 & 2 \\ \hline
51 & 15:16:21 & CH1-LONG & 1531-1307 & 527.258 & 2 \\
 &  & CH2-LONG & 1000-855 & 527.258 & 2 \\
 &  & CH3-LONG & 649-556 & 316.355 & 2 \\
 &  & CH4-LONG & 413-358 & 105.452 & 2 \\ \hline
140.5 & 17:46:43 & CH1-SHORT & 2040-1742 & 527.258 & 2 \\
 &  & CH2-SHORT & 1331-1140 & 527.258 & 2 \\
 &  & CH3-SHORT & 866-742 & 316.355 & 2 \\
 &  & CH4-SHORT & 565-477 & 105.452 & 2 \\ \hline
130.5 & 17:30:04 & CH1-MEDIUM & 1766-1508 & 527.258 & 2 \\
 &  & CH2-MEDIUM & 1153-987 & 527.258 & 2 \\
 &  & CH3-MEDIUM & 750-642 & 316.355 & 2 \\
 &  & CH4-MEDIUM & 483-408 & 105.452 & 2 \\ \hline
120 & 17:13:17 & CH1-LONG & 1531-1307 & 527.258 & 2 \\
 &  & CH2-LONG & 1000-855 & 527.258 & 2 \\
 &  & CH3-LONG & 649-556 & 316.355 & 2 \\
 &  & CH4-LONG & 413-358 & 105.452 & 2 \\ 
\end{tabular}%
}
\label{table1}
\end{table}

\begin{figure}[t]
    \centering
    \includegraphics[width=\textwidth]{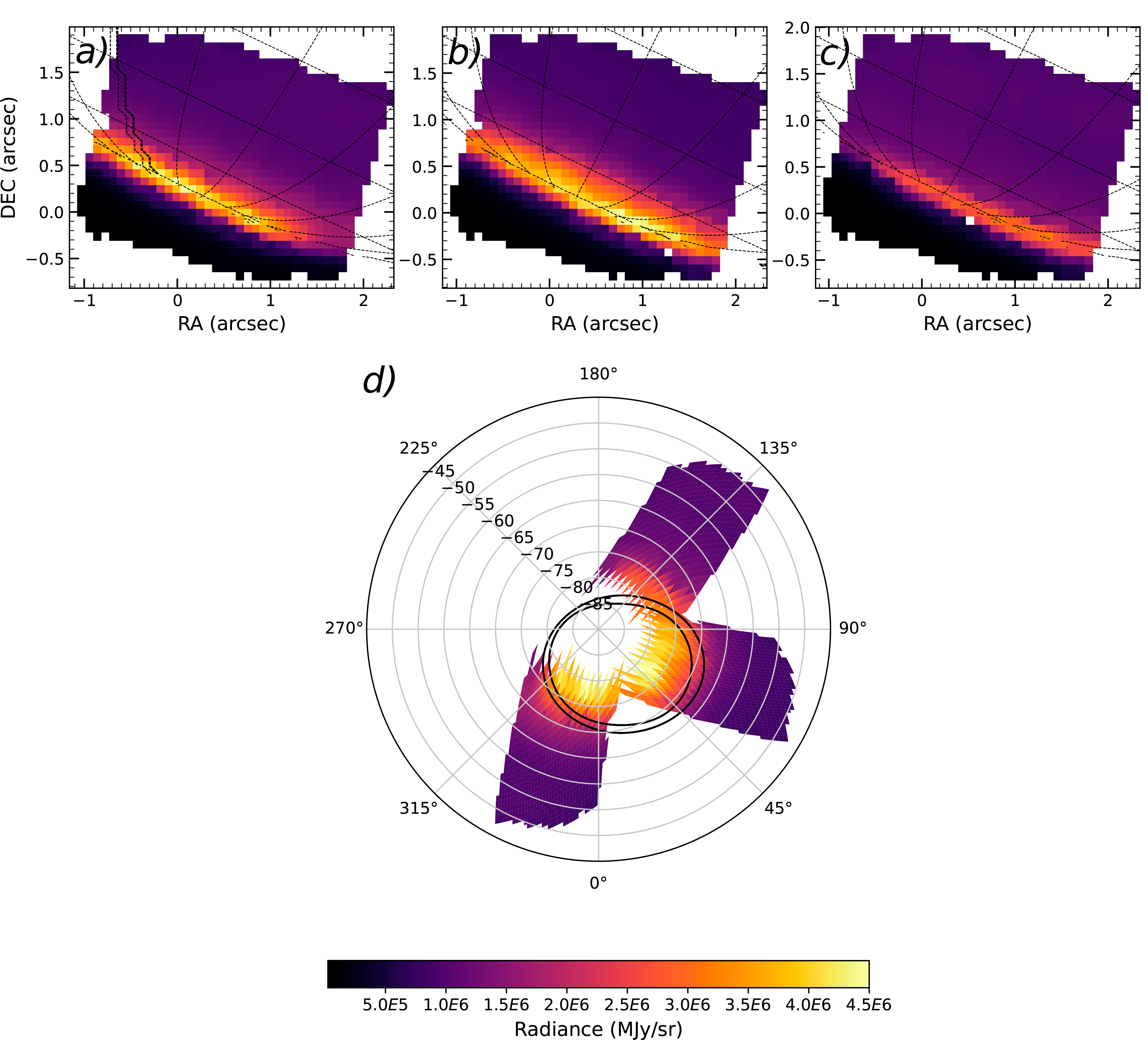}
    \caption{Radiance maps in MJy/sr for the $\nu_4$ CH$_4$ Q-branch at 1306~cm$^{-1}$ in two different projections. a) Tile~1, centered at 340$^{\circ}$W, in orthocentric projection (coordinates relative to the south pole). b) Tile~2, centered at 70$^{\circ}$W, in orthocentric projection. c) Tile~3, centered at 140$^{\circ}$W, in orthocentric projection. d) Polar projection of the three tiles from 45$^{\circ}$S to 90$^{\circ}$S with the black solid lines showing the statistical position of the inner and outer borders of the auroral oval \cite{Bonfond2017May}.}
    \label{radiance}
\end{figure}

\begin{figure}[t]
    \centering
    \includegraphics[width=\textwidth]{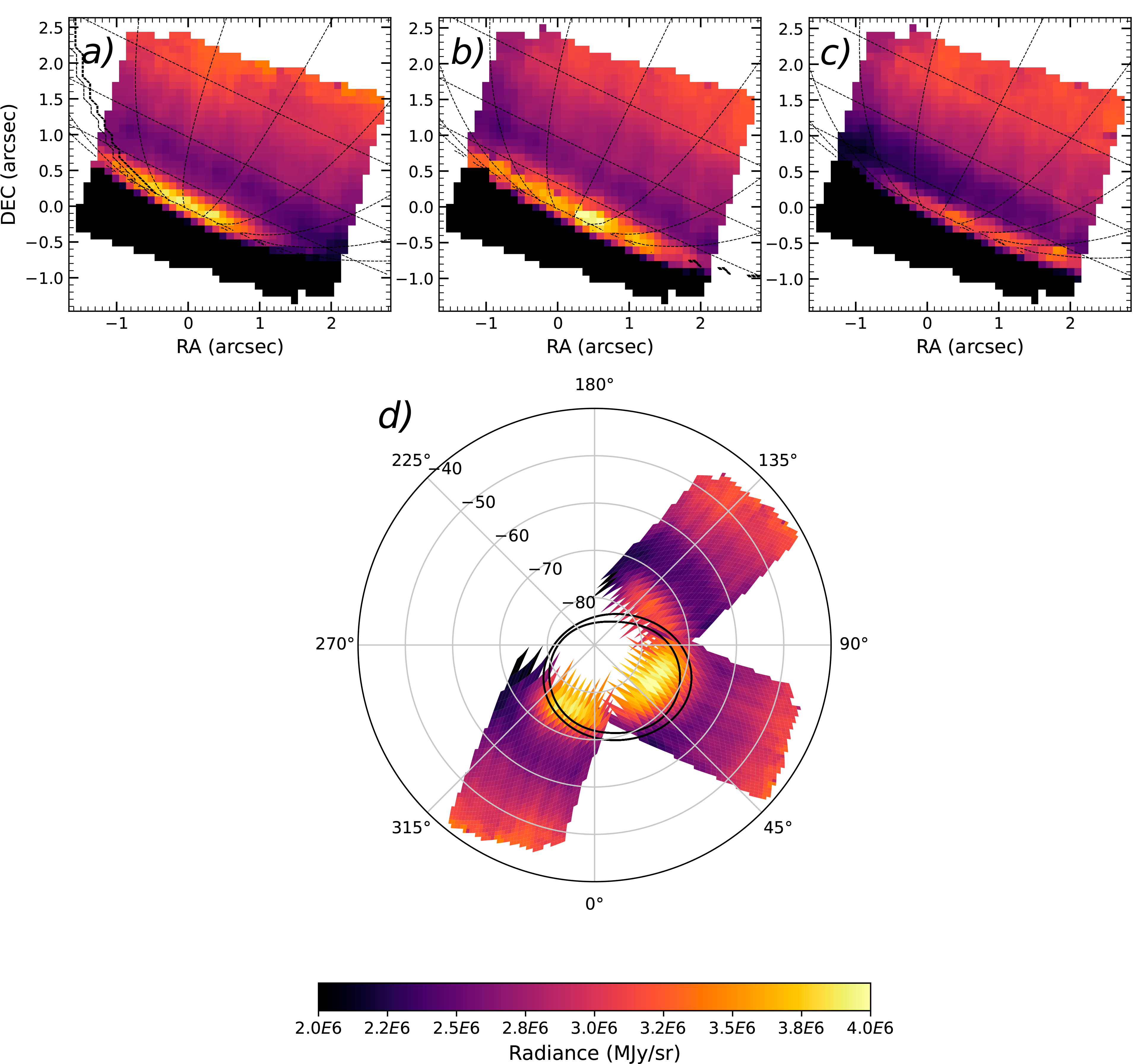}
    \caption{Same as Fig.~\ref{radiance} in the $\nu_5$ C$_2$H$_2$ band at 694 cm$^{-1}$. Polar projection from 40$^{\circ}$S to 90$^{\circ}$S }
    \label{radiance2}
\end{figure}

\begin{figure}[t]
    \centering
    \includegraphics[width=\textwidth]{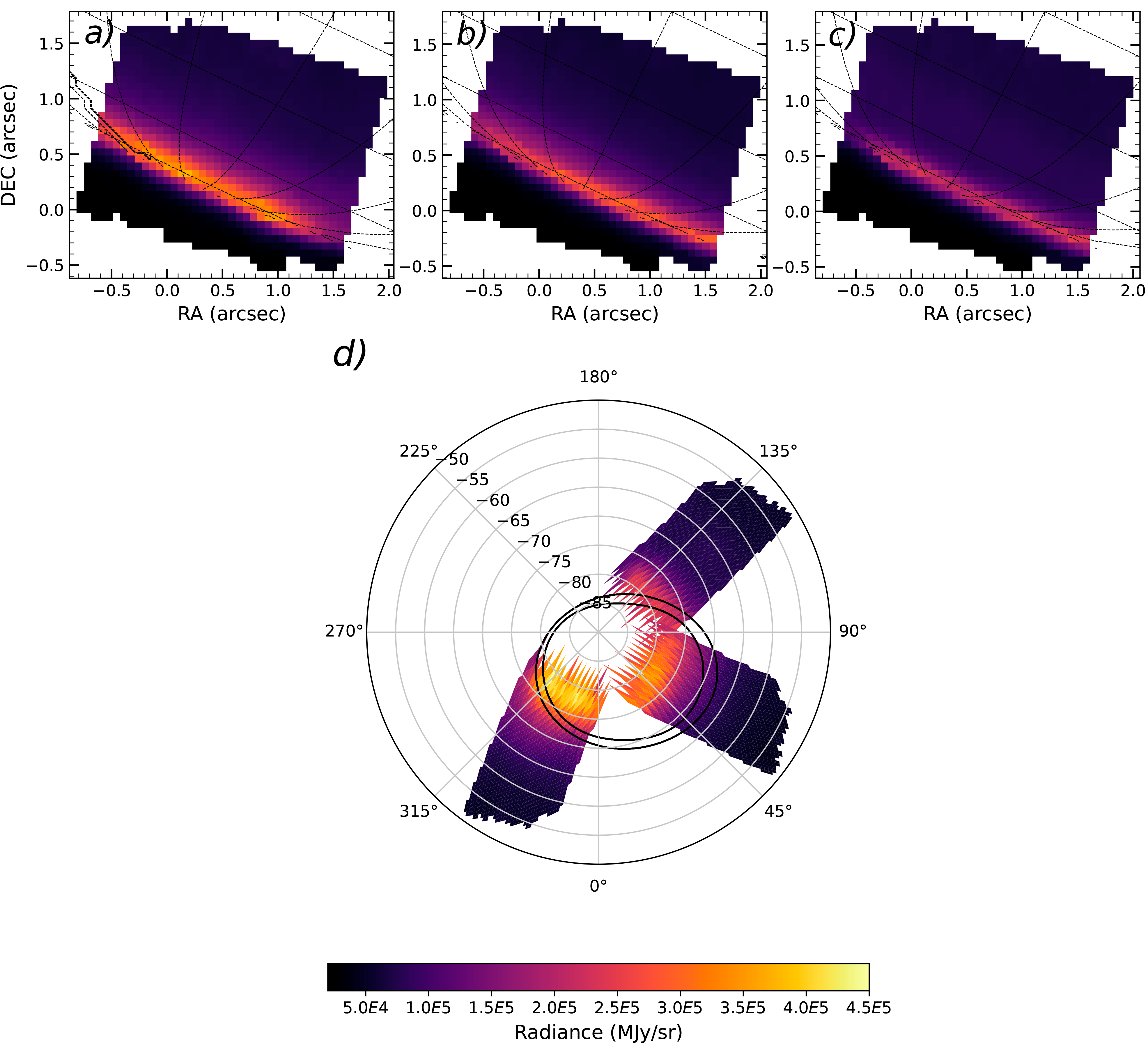}
    \caption{Same as Fig.~\ref{radiance} in the $\nu_8$ C$_2$H$_6$ band at 1520 cm$^{-1}$. Polar projection, poleward of 50$^{\circ}$S.}
    \label{radiance3}
\end{figure}

In our observations, we mapped the SPR with a mosaic of 3 tiles centered at different longitudes. Each tile was observed using a 2-point dither pattern and a 527.258 s exposure time per sub-band for 5 groups. Our observations covered latitudes poleward of 50$^{\circ}$S and the FOV for each observation was centered at 340$^{\circ}$W, 70$^{\circ}$W and 140$^{\circ}$W (System III). The detector readout was performed in FASTR mode to account for the brightness of our target, since Jupiter would saturate slower readout modes. Fig.~\ref{radiance} shows the three navigated cubes from channel 2-SHORT at 1306 cm$^{-1}$ (Q branch emission of the CH$_4$  $\nu_4$ band). Figs. \ref{radiance2} and \ref{radiance3} display the three navigated cubes from channel 3-MEDIUM at 714 cm$^{-1}$ and 1-MEDIUM at 1515 cm$^{-1}$, probing the emission of C$_2$H$_2$ and C$_2$H$_6$ respectively.

MIRI and the other instruments on board JWST are based on a non-destructive up-the-ramp readout \cite{MIRIfocal}, which means that the total integrations of our observation is divided into a user-specified number of groups (5 in our case). Each group is then downlinked to the ground and included in JWST archived data.
Our dataset was processed using the 1.11.3 version of the JWST pipeline \cite{pipeline_bushouse}, and the CRDS (Calibration References Data System) file $jwst\_1119.pmap$.  This pipeline consists of 3 distinct steps or 'stages'. The objective of 'Stage 1' is to apply corrections at the detector level, i.e., subtraction of the dark current, subtraction of the detector superbias (This step removes the fixed detector bias from a science data set by subtracting a superbias reference image (https://jwst-pipeline.readthedocs.io/en/latest/jwst/superbias/description.html)), as well as ramp fitting by means of a linear fit of different groups within an integration. This process allows for the calculation of counts per second for each pixel in the detector image. In this stage, pixels can be flagged as saturated if the linear fit is not optimum (see Section \ref{sec_saturation}). Stage 2 applies instrument-level corrections. This includes flat field correction, photometric corrections, background subtraction, and conversion of the count rates into physical units (MJy/sr). Finally, 'Stage 3' combines the calibrated products from the previous stage, and converts the detector images into hyperspectral cubes.

In addition to this standard pipeline process, a series of specific processes required by our observations have been carried out to correct artifacts that we found affected our dataset. In the following sections, we will show these artifacts and explain the strategy followed to correct them. 

\subsection{Saturation}
\label{sec_saturation}

The high sensitivity of the instruments on board JWST becomes problematic when working with dataset of bright bodies. This is specifically the case for our MIRI observations. In the MIRI wavenumber range, Jupiter's brightness temperature is never lower than $\sim$120K. For this reason, our program was scheduled to be carried out with 16 integrations per exposure and 5 groups per integration, and the detector readout set to the FASTR mode. This configuration allowed us to limit the occurrence of saturated pixels at wavenumbers shortwards of 1000 cm$^{-1}$.

This saturation problem can be solved thanks to the readout mode chosen for the JWST detectors \cite{MIRIfocal}. This readout process, and the availability of all the groups, allows us to manually reduce the effective integration time by reducing the number of groups used in the data reduction procedure. As a result, we can effectively mitigate the saturation problem.

\begin{figure}[t]
    \centering
    \includegraphics[width=\textwidth]{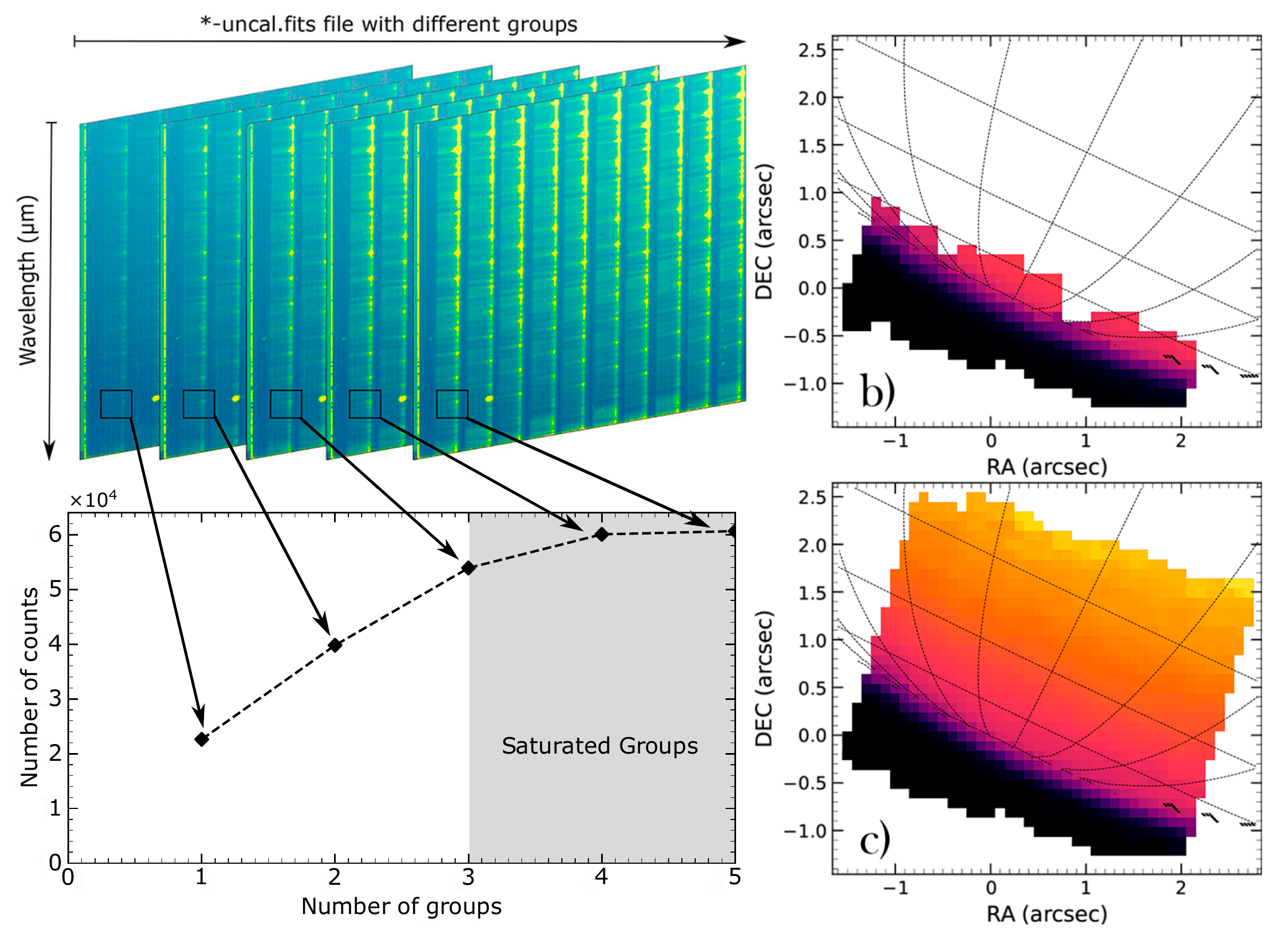}
    \caption{a) Example of saturated spaxel in the stacked uncalibrated files (top) and the saturation of the spaxels due to the loss of linearity (bottom). b) Radiance map at 746 cm$^{-1}$ obtained using the standard version of the pipeline for a wavenumber of 746 cm$^{-1}$. c) Radiance map at 746 cm$^{-1}$ obtained using our desaturation methodology. }
    \label{desat}
\end{figure}

Fig.~\ref{desat} a) shows how this up-the-ramp readout works. The linearity of the count readout, with respect to the number of groups, serves as an indicator of the readout quality. When the number of counts reaches a certain threshold value \cite{Glasse2015Jul}, the linearity is lost, and the pipeline flags this readout as saturated.

For each detector readout, we have created five uncalibrated files (prior to running the pipeline), each including a specific number of groups, respectively the 1st, 1st and 2nd groups, 1--3 groups, 1--4 groups, and the 5 groups of the detector ramps. These uncalibrated files were subsequently processed through the regular three stages of the pipeline to produce five new calibrated hyperspectral cubes. To obtain the final hyperspectral cube, we combined the five different calibrated hyperspectral cubes. For each wavenumber and spaxel (spatial pixel in a reconstructed data cube stores the spectrum associated to a spatial element projected on the sky), we assigned the radiance from the hyperspectral cube created with the highest number of groups that were not saturated, as in \cite{Fletcher2023Sepsaturn} and \cite{King2023Oct}. This approach was designed to maximize the signal-to-noise ratio. While this method allowed us to recover spectral information shortwards of 1000 cm$^{-1}$, it was unable to completely desaturate some features, such as the C$_2$H$_2$ $\nu_5$ Q-branch at 730 cm$^{-1}$, and the C$_2$H$_6$ $\nu_9$ band centered at 822 cm$^{-1}$. We also stress that using a smaller number of groups makes it more challenging to reject cosmic rays. We must keep in mind that some spaxels and wavenumbers may be statistically more affected by cosmic rays than others. Units were finally converted from MJy/sr to W cm$^{-2}$ sr$^{-1}$ / cm$^{-1}$.

\subsection{Spectral calibration}

The spectral resolution and high signal-to-noise ratio (SNR) of MIRI-MRS allowed us to identify a residual error in the pipeline wavenumber calibration process. For certain regions of the detector, the flat field correction applied by the version 1.11.3 of the pipeline exhibited deviations from the expected wavenumber calibration, leading to a small residual wavenumber deviation ($\sim0.25$~cm$^{-1}$) in certain spaxels with respect to others. As a result, there appeared to be spatial striping, as different spaxels sampled different wavenumber offsets from the center of a given emission line, resulting in varying radiances. To address this issue, a new spectral calibration using observations of Jupiter GTO $\#$1246 and Saturn GTO $\#$1247 was used to improve the quality of the dataset used in this work. This new calibration step was developed following the procedure presented in \cite{Argyriou2023Mar}. These authors used the Jupiter and Saturn spectra (along with spectra from other programs) and compared them with synthetic spectra generated using the NEMESIS radiative transfer code \cite{Irwin2008Apr}. This comparison allowed them to determine the residual wavenumber shift as a function of wavenumber in the range of 2000 to 600 cm$^{-1}$ and to propose a specific correction. This calibration step was already been validated by \cite{Fletcher2023Sepsaturn} against the MIRI-MRS spectra of Saturn.

The spectra shortwards of 600 cm$^{-1}$ are also affected by partial and total saturation. Even with only 1 group, the spectra, which should follow the shape of the black-body emission of Jupiter ($\sim$120K) in this range, presents a series of wavelike features in addition to a saw-tooth noise always located at the same spectral positions. These features prevented us from using this spectral range in our analysis with the current state of the pipeline.

\subsection{Selection of the spectral regions}

To investigate the thermal structure of Jupiter's stratosphere, we inverted spectra covering the CH$_4$ $\nu_4$ band, from 1240 to 1330 cm$^{-1}$ (channel 2-SHORT), as in \cite{Fouchettemp}. This spectral region allows us to probe the atmosphere in the pressure range between 0.01 and 20 mbar. We excluded the H$_2$ S(1) line from our analysis, since the spectra at wavenumbers below 600 cm$^{-1}$ are saturated, especially in the auroral region (see Sect.~\ref{sec_saturation}). 

We also used the 1510 -- 1570 cm$^{-1}$ spectral range (channel 1-MEDIUM), where we observed a non-negligible contribution from the C$_2$H$_6$ $\nu_{8}$ band, to retrieve the volume mixing ratio (VMR) of this hydrocarbon (Section \ref{sec_res_hydroc}). Indeed, we have excluded the $\nu_{9}$ band centered at 822 cm$^{-1}$, which is commonly used in the literature to retrieve the C$_2$H$_6$ abundance because in our MIRI-MRS dataset it is affected by partial saturation. Neighboring the $\nu_8$ C$_2$H$_6$ band, CH$_4$ emission lines from the $\nu_2$ band are clearly visible. This band was used to retrieve the temperature in the pressure region from 0.1 -- 30 mbar. Unlike the $\nu_4$ band, the $\nu_2$ band is a forbidden band with a weak Einstein coefficient and is not sensitive to higher altitudes.

The 680 -- 760 cm$^{-1}$ (channel 3-SHORT and MEDIUM) spectral range was used to retrieve the abundance of C$_2$H$_2$ through emissions in its $\nu_5$ fundamental and harmonics bands. However, we excluded the $\nu_5$ band Q-branch centered at 730 cm$^{-1}$ from our analysis. This particular branch remained saturated even at the lowest number of groups, especially in the auroral region. We also excluded the spectra between 695 and 705 cm$^{-1}$, as it is affected by aerosol spectral features. It is important to note that the spectral resolution in the MRS mode varies across the different spectral regions. For the CH$_4$ $\nu_2$, C$_2$H$_6$, and CH$_4$ $\nu_4$ spectra, the spectral resolution is approximately 3700. On the other hand, for the C$_2$H$_2$ spectra, the spectral resolution is approximately 2400.

\begin{figure}[ht]
    \centering
    \includegraphics[width=\textwidth]{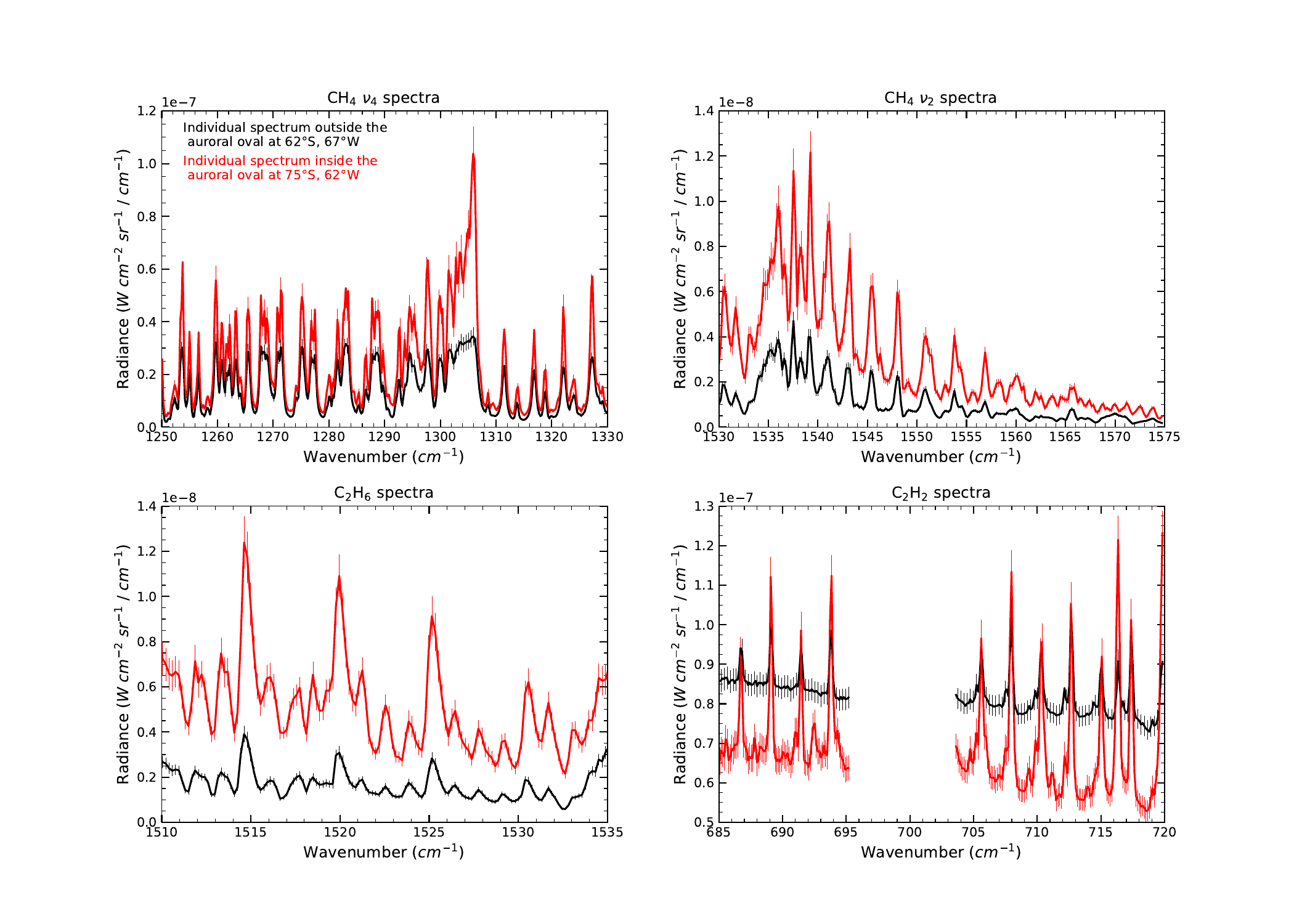}
    \caption{Example of the spectra used in this work for a spectra located within (75$^{\circ}$S, 62$^{\circ}$W) and outside (62$^{\circ}$S, 67$^{\circ}$W) the auroral oval in red and black, respectively. Top left: Spectra from the $\nu_4$ band of methane, used to retrieve the stratospheric temperature and the homopause height. Top right: Spectra from the $\nu_2$ band of methane, used to retrieve the stratospheric temperature. Bottom left: C$_2$H$_6$ spectra used to retrieve its abundance profile. Bottom right: C$_2$H$_2$ spectra used to retrieve its abundance profile. The error bars are shown as thin vertical lines. Note the different y-axis scales.}
    \label{all_spe}
\end{figure}

In summary, Fig.~\ref{all_spe} presents a comparison of the three spectral ranges used in this study. The plot displays spectra obtained both inside and outside the auroral region for each range. 

\subsection{SNR estimations}\label{SecSNR}
 
The error matrix associated with the MRS hyperspectral cubes exhibits very low values, with signal-to-noise ratios (SNRs) reaching as high as 5000 in some cases. However, after the desaturation process, this SNR is scaled by the fraction of groups remaining in our $'*uncal.fits'$ files. As a result, our final SNRs are overall smaller than those given by the pipeline.

When calculating the SNRs, the pipeline considers only the photon noise, readout noise, and detector noise components. However, for a bright target such as Jupiter, this approach underestimates the total noise present in the observations. Other noise sources, such as the noise due to calibration uncertainties, raising from the wavenumber calibration (striping), and other instrumental artifacts can contribute more significantly to the overall noise level. Therefore, the pipeline SNR estimation may not accurately reflect the true noise affecting our dataset.
To address this issue, we have estimated the noise level present in the H$_2$-He-CH$_4$ continuum in channel 3-MEDIUM (645 -- 665 cm$^{-1}$) to obtain an associated SNR. From this value, we scaled the SNR value of the hyperspectral cubes until it is close to the SNR obtained in the continuum. In general, it was necessary to multiply the noise level given by the pipeline by a factor of 50 -- 70, so the noise level actually reflects the true quality of our observations. After this correction, the resulting SNR has a value of $\sim$ 100 for the spectra in channel 2, and of $\sim$ 70 for the spectra in channel 3.

\subsection{Spatial registration} \label{navigation}

We projected each hyperspectral cube onto the disk of Jupiter using the JWST SPICE kernels (Version from July 2023) for the spatial registration. For each spaxel, we calculated planetographic and planetocentric latitude and longitude information, as well as the emission and incidence angles, and the distance to the limb of the planet, based in the pointing coordinates provided by the metadata of the hyperspectral cubes.

Due to the significant time gaps between sub-bands, we processed each sub-band individually for each channel. Consequently, for each tile of our $3\times1$ mosaic, we performed navigation separately for the 12 hyperspectral cubes corresponding to each sub-band, as each one of them also has different pixel size and the pointing slightly changes between them. As Jupiter rotated by up to 10 degrees between sub-band observations, it was not feasible to use the same navigation data for all sub-bands within the same channel. Moreover, we needed to account for minor variations in telescope pointing between sub-bands, especially when analyzing spectra near the limb of the planet. The spatial registration code takes into account several parameters, including the telescope pointing, observation time, and the spaxel size specific to each channel and sub-band. For instance, at 70$^{\circ}$S, one spaxel projects onto 1$^{\circ}$ of longitude on the planet for wavenumbers centered at 1500 and 1300 cm$^{-1}$, while for wavenumbers centered at 700 cm$^{-1}$ the spatial resolution decreases, so that one spaxel at 70$^{\circ}$S projects as 2$^{\circ}$ of longitude on the planet.

\section{Data analysis}\label{sec_data}

In the mid-infrared spectral ranges that we have analyzed, the temperature and molecular abundances are the main drivers in shaping the intensity of the emission lines. To reproduce and invert these emission spectra, we used a line-by-line radiative transfer code able to generate synthesized spectra from a given temperature vertical profile and prescribed molecular abundance profiles. We assume that the emission is in Local Thermodynamic Equilibrium (LTE) at all pressure levels and discuss the associated limitations in Sec.~\ref{sec_discussion}. We divide the atmosphere in 361 layers, equally spaced in a logarithmic scale of pressure from 10 to $10^{-8}$ bar. The code takes into account the latitudinal and vertical variations of gravity using the latest measurements of Jupiter's gravity fields and rotation rate obtained by the Juno spacecraft \cite{Iess2018Mar}. Since our observations of the SPR encompassed some grazing angles, the length of the light path was calculated in spherical geometry. To do that, for each spaxel, we calculate the cosine of the local emission angle ($\mu (z)$) for each layer, following:

\begin{equation}
    \mu (z)= \sqrt{1-\left(\frac{R}{R+z} \sqrt{1-\mu_e^2}\right)^2}  ,
    \label{eq_sphere}
\end{equation}

where $\mu_e$ is the cosine of the emission angle for a height of 0 km (corresponding to the 1-bar level). $R$ is the local radius of the planet at a given latitude, and $z$ is the altitude of each layer with respect to the 1-bar pressure level.

Our model takes into account opacities using the HITRAN 2020 database \cite{Gordon2022Jan}. The model includes the opacities of CH$_4$, CH$_3$D, NH$_3$, PH$_3$, C$_2$H$_2$, and C$_2$H$_6$. Furthermore, we take into account the collision-induced continuum of H$_2$-He-CH$_4$ in the same way as proposed by \cite{Borysow1985, Borysow1988}.

The deep volume mixing ratio (VMR) for methane is set to $2.04\times 10^{-3}$, as measured in-situ by the Galileo probe \cite{Wong2004Sep}. For CH$_3$D, the deep VMR is set to $1.4\times 10^{-7}$, consistent with the analysis presented in \cite{Lellouch2001}. For hydrocarbons, initial \textit{a priori} profiles have been taken from the photochemical model of \cite{Moses2017}. This model includes the complete chemical pathway presented in \cite{Moses2005} for the hydrocarbon chemical reactions triggered by photolysis. It has also been updated with ablation processes, that include the injection of exogenic species into the atmospheres from micrometeorites. The altitude of the homopause can be changed by varying the gradient of the eddy diffusion coefficient, as shown in \cite{Sinclairhomopause}.

\subsection{Inversion algorithm}\label{sec_inversion}

The retrieval of the vertical temperature or chemical abundance profiles from spectroscopic observations constitutes a challenge due to the degeneracy of possible solutions, which makes the retrieval of the atmospheric structure an ill-posed problem \cite{Rodgers2000Jul}. In this work, we used a regularized retrieval algorithm detailed in \cite{Conrath1998} and used in several studies, such as \cite{Guerlet2009Sep} and \cite{Fouchettemp} for Saturn. Starting from an \textit{a priori} profile, this method inverts a posterior profile that provides the best fit to the observed spectra, smoothly departing from the \textit{a priori} profile in pressure ranges where the information from the spectra dominates. Thus, starting from \textit{a priori} profiles of temperature and abundances, the thermal and chemical profiles retrieved will remain close to the \textit{a priori} profiles at pressure levels where there is little information content, while at pressure levels probed by the observations, the retrieved profile will depart from the \textit{a priori}. This process helps to mitigate the ill-posed nature of the inversion and provides more reliable atmospheric structure estimates.

Our algorithm assumes that the radiance can be linearized as a function of the model variables (temperature and abundance profiles) as follows.

\begin{equation}
    \Delta I_i = \sum_{j=1}^{n}\frac{\partial I_i}{\partial x_{1,j}}\Delta x_{1,j} + \frac{\partial I_i}{\partial x_{2,j}}\Delta x_{2,j}
    \label{jacobian}
\end{equation}

where $I$ is the radiance at a specific wavenumber ($\tilde{\nu_i}$), and $x$ the model variables, in our case the temperature $T_j$ ($x_1$) and the natural logarithm of the abundances $ln(q_j)$ ($x_2$). These variables are vectors, with the index $j$ denoting the pressure layer. $\Delta x_j$ represents the variation at a particular pressure level that will be added to the profile during a given iteration $n$ to generate the reference profile for the subsequent iteration $n+1$, from which the synthetic spectrum will be calculated. In Section \ref{sec_res_temp} we inverted the stratospheric temperature only ($x_1$), while in Section \ref{sec_hydroc_abund} we inverted the tropospheric temperature to fit the continuum ($x_1$) and the abundance of the hydrocarbon analyzed ($x_2$).

For clarity, this equation can be written in a simpler format, where we denote the derivative matrices as $K_1$ and $K_2$ with respect to the corresponding parameters $x_1$ and $x_2$, so that:

\begin{equation}
    \Delta I_i = K_1 \Delta x_{1} + K_2\Delta x_{2}
    \label{change_delta}
\end{equation}

The formal solution to this ill-posed problem for the two variables $x_1$ and $x_2$ can be written as:

\begin{equation}
    \Delta x_1 = U \Delta I \textrm{ with } U = \alpha S K_1^T(\alpha K_1 S K_1^T + \beta K_2 S K_2^T + E^2)^{-1} 
    \label{change_Tq}
\end{equation}
\begin{equation}
    \Delta x_2 = V \Delta I \textrm{ with } V = \beta S K_2^T(\alpha K_1 S K_1^T + \beta K_2 S K_2^T + E^2)^{-1}
    \label{change_Tq2}
\end{equation}

where $S$ is the covariance matrix that smooths the variations in the variables by a given vertical length (given in scale heights). The matrix $E$ is the covariance matrix containing the measurement errors. In our case, the matrix $E$ is supposed to be diagonal. The parameters $\alpha$ and $\beta$ are scalar weight values that establish the balance between the \textit{a priori} values and the information coming from the spectra. \cite{GuerletThesis} found that these parameters ($\alpha$ and $\beta$) are optimal when their values are set to equal the traces of the $E^2$ with the $\alpha K_1SK_1^T$ and $\beta K_2SK_2^T$ matrices.

The algorithm proceeds with a series of iterations, modifying the vertical profile at each step. By solving Eq.\ref{change_Tq}, we obtain the variations of $\Delta x_1 = \Delta T$ and $\Delta x_2 = \Delta ln(q)$ that are added to the previous vertical profile to generate new profiles that will be used as input for the radiative transfer code in the next iteration. Thus, the new value of the vector $T$ would be $T_0+\Delta T$ and the abundance $q$ would be $q_0 \times (1+ e^{\Delta ln(q)})$. In addition to the input variables, the altitude grid changes as it depends on temperature, as well as the functional derivatives. The convergence of these iterations is governed by the quantity $\chi^2$, an indicator of the goodness of fit, which compares the radiance of the synthetic spectrum generated by our algorithm with the radiance measured by MIRI-MRS (Equation \ref{chi2eq}). Iterations continue until a convergence criterion is reached, when the relative change in $\chi^2$ between two successive iterations is less than or equal to 1\%.

\begin{equation}
    \chi^2 = \sum \left ( \frac{\Delta I_i}{E_i} \right )^2
    \label{chi2eq}
\end{equation}

To estimate the information content of the retrieval, we used the averaging kernel matrix $A = UK$. Each of the rows of the matrix $A$ represents the ratio between the relative weight of the measurement information, and the information from the \textit{a priori} profile itself. Thus, as long as the peak of the function of each row ($a_j^T$) reaches a significant value at the corresponding same pressure level $p_{j}$, it means that the temperature or abundance information at that pressure level comes mainly from the measurement. Therefore, matrix A can be used to analyze the range of pressures probed by our measurements. Furthermore, we can quantify the number of independent pressure levels to which we are sensitive, also known as the degrees of freedom of the signal (d). This can be calculated using the expression:

\begin{equation}
    d = Tr(A)
\end{equation}

\subsection{Information content for the inversion of the thermal structure}\label{SecInformationContent}

The comparison between different spectra displayed in Fig.~\ref{all_spe} clearly reveals the large difference in radiance between a spectrum obtained inside and outside the auroral oval. In fact, the spectra obtained in the polar auroral region cannot be satisfactorily fitted with the models assuming low CH$_4$ abundances at higher altitudes, which suggests an upward shift of the homopause level.

\begin{figure}[ht]
    \centering
    \includegraphics[width=\textwidth]{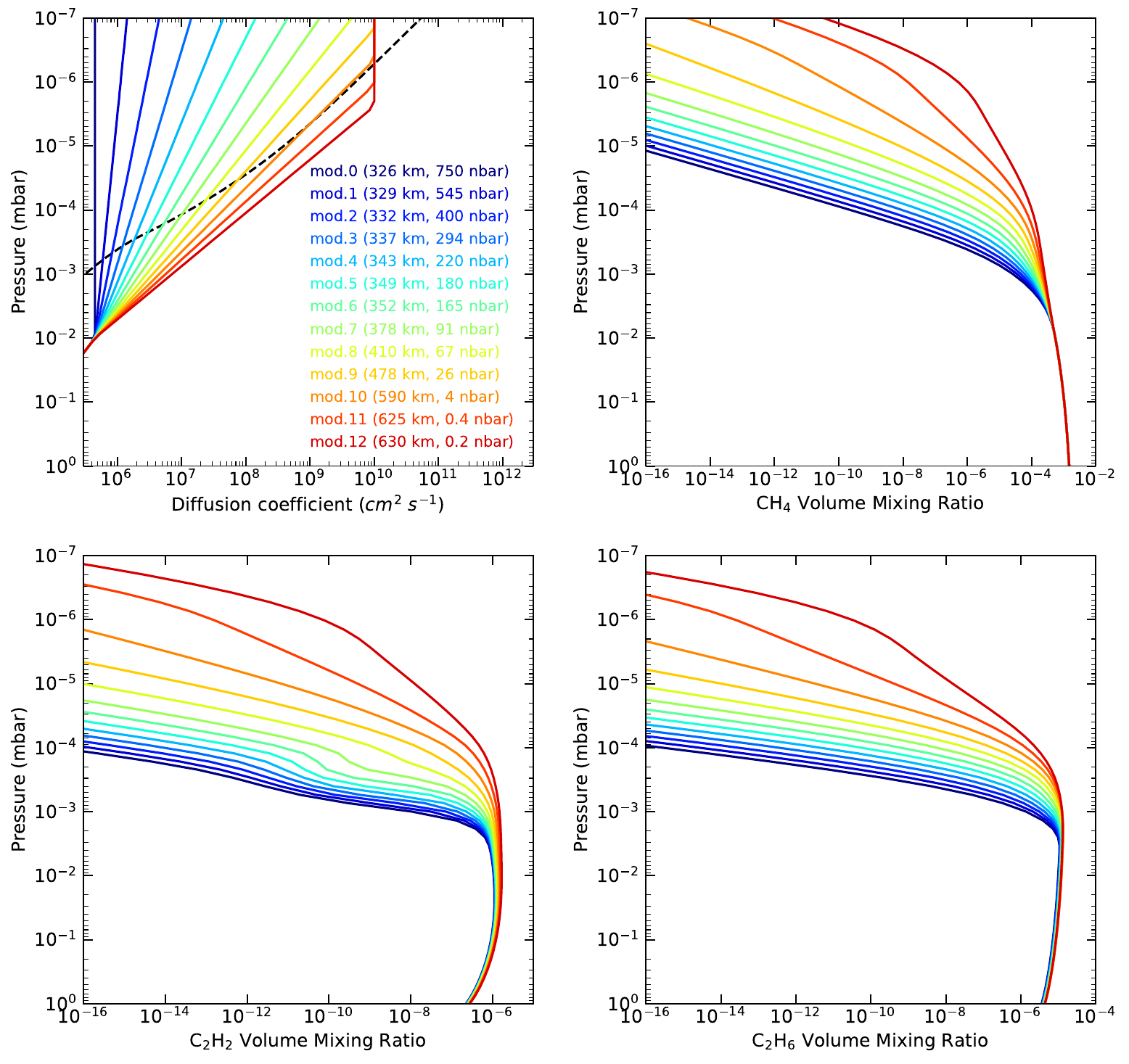}
    \caption{Eddy diffusion coefficients (black dashed line corresponding to the molecular diffusion coefficient) and vertical profiles of CH$_4$, C$_2$H$_2$ and C$_2$H$_6$ for the 13 homopause models used, varying the homopause altitude with respect to the 1 bar pressure level, indicated in the top-left panel for each model. The color scheme in this figure is adopted throughout the paper.}
    \label{ch4_profiles}
\end{figure}

While it is possible to retrieve simultaneously the temperature and the CH$_4$ vertical profiles, the degeneracy between these two variables makes the inversion unstable. Following the approach proposed by \cite{Sinclairhomopause}, we found a more stable solution to only retrieve the temperature profile ($x_1$ from Eq.~\ref{jacobian}) using thirteen different CH$_4$ vertical profiles. These profiles remain fixed throughout the inversion process. For each spaxel in our dataset, we compare the fits obtained for every CH$_4$ profile and determine the CH$_4$ vertical profile that yields the best fit. We then adopt the associated inverted temperature profile as our solution temperature profile for the given spaxel. Each CH$_4$ profile is the result of the use of different eddy diffusion coefficients. We can assume then that each profile corresponds to a CH$_4$ homopause height, as the homopause is defined as the pressure at which the molecular diffusion coefficient equals the eddy diffusion coefficient. Fig.~\ref{ch4_profiles} shows the different profiles used for our determination of the homopause location. As expected, we can clearly see an increase in the abundance of methane at higher altitudes associated with the upward displacement of the homopause. Model $\#$0 corresponds to the lowest homopause height ($\sim$326 km or 750 nbar), while model $\#$12 corresponds to the highest homopause height ($\sim$630 km or 0.2 nbar).

\begin{figure}[ht]
    \centering
    \includegraphics[width=\textwidth]{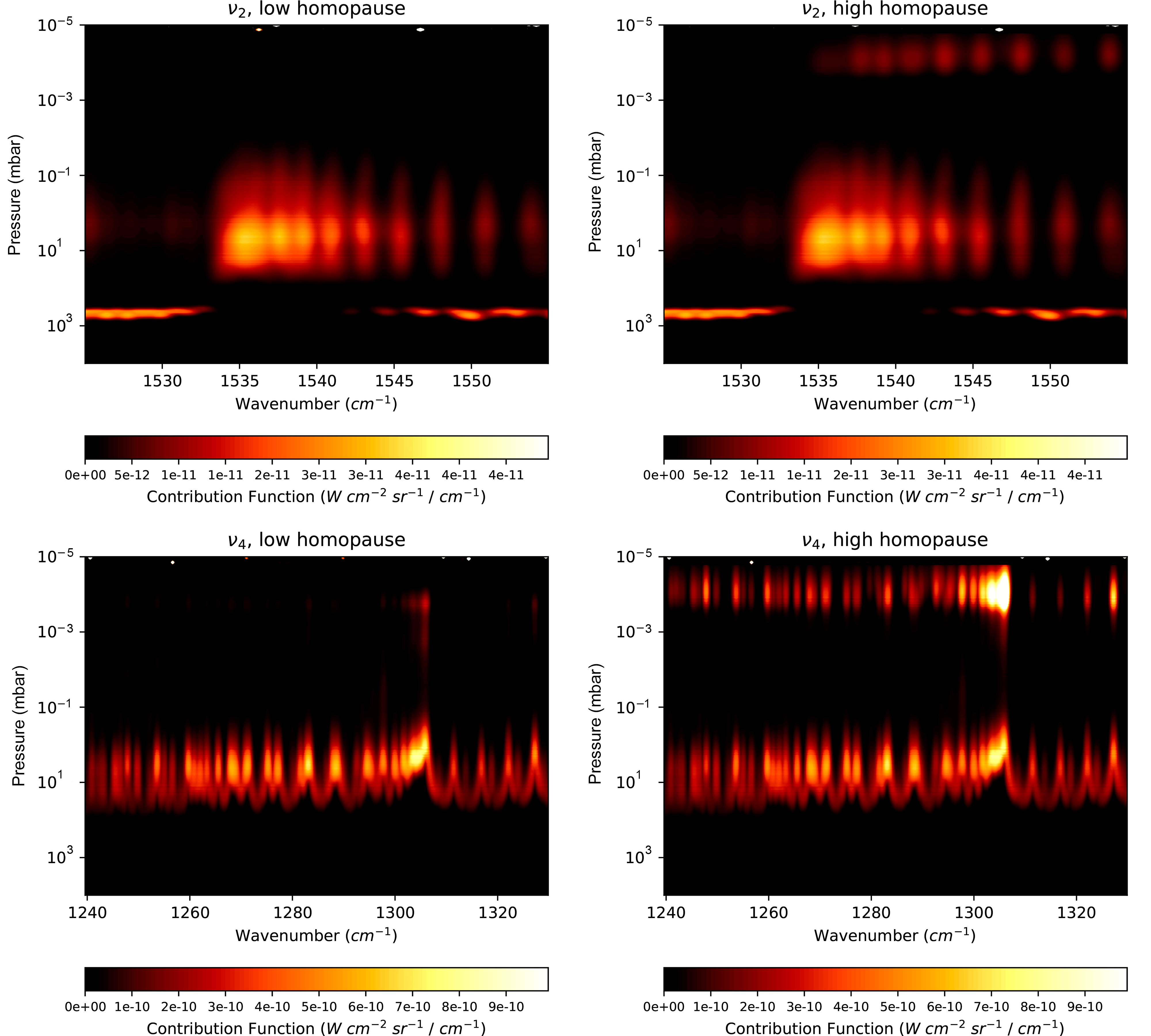}
    \caption{Contribution functions within the CH$_4$ $\nu_2$ band (top row) and within the CH$_4$ $\nu_4$ band (lower row) for two different methane vertical profiles: profile 6 (left column) and profile 10 (right column) from Fig.~\ref{ch4_profiles}. The thermal profile used for this calculation corresponds to the temperature profile used in \cite{Moses2005}.}
    \label{contfunc}
\end{figure}

Given the spectral information provided within the MIRI-MRS spectral range, the thermal profile can be inverted in three different ways: using the $\nu_2$ band of CH$_4$ only, using the $\nu_4$ band of CH$_4$ only, or using both bands simultaneously. Fig.~\ref{contfunc} illustrates the contribution functions for both bands, considering two different CH$_4$ VMR profiles characteristic of a low and a high homopause, respectively (models 6 and 10 from Fig.~\ref{ch4_profiles} and models 3 and 7 in \cite{Sinclairhomopause}) using the thermal profile used in \cite{Moses2005}. We see that the $\nu_4$ band provides information near 1-20 mbar, but also at higher pressure levels ($\sim$ 0.1 $\mu$bar) for a high homopause conditions. Complementary, the $\nu_2$ band has the largest information content between 20 and 0.1 mbar, with a significant contribution near 800 mbar, and also an increase at higher altitude for a high homopause although not as large as the $\nu_4$ band. Although the $\nu_2$ band exhibits an increase in information at these high altitudes for a model with the homopause located at higher altitudes, this increase is not as pronounced as in the $\nu_4$ band. 
This indicates a much larger information content in the $\nu_4$ band. Furthermore, the advantage of the $\nu_4$ spectral range is that it does not feature emissions from other molecules, such as C$_2$H$_6$ affecting CH$_4$ $\nu_2$ spectral range, which forces us to perform a simultaneous inversion of ethane abundance and temperature. These factors make the $\nu_4$ band more useful than the $\nu_2$ band to obtain information on the height of the homopause, given the low sensitivity of this band to atmospheric parameters at high altitudes, around the $\mu$bar pressure level (see the supplementary materials for more information on the $\nu_2$ band). We hence decided to only use the CH$_4$ $\nu_4$ band for the retrievals of the stratospheric temperatures.

\begin{figure}[ht]
    \centering
    \includegraphics[width=0.8\textwidth]{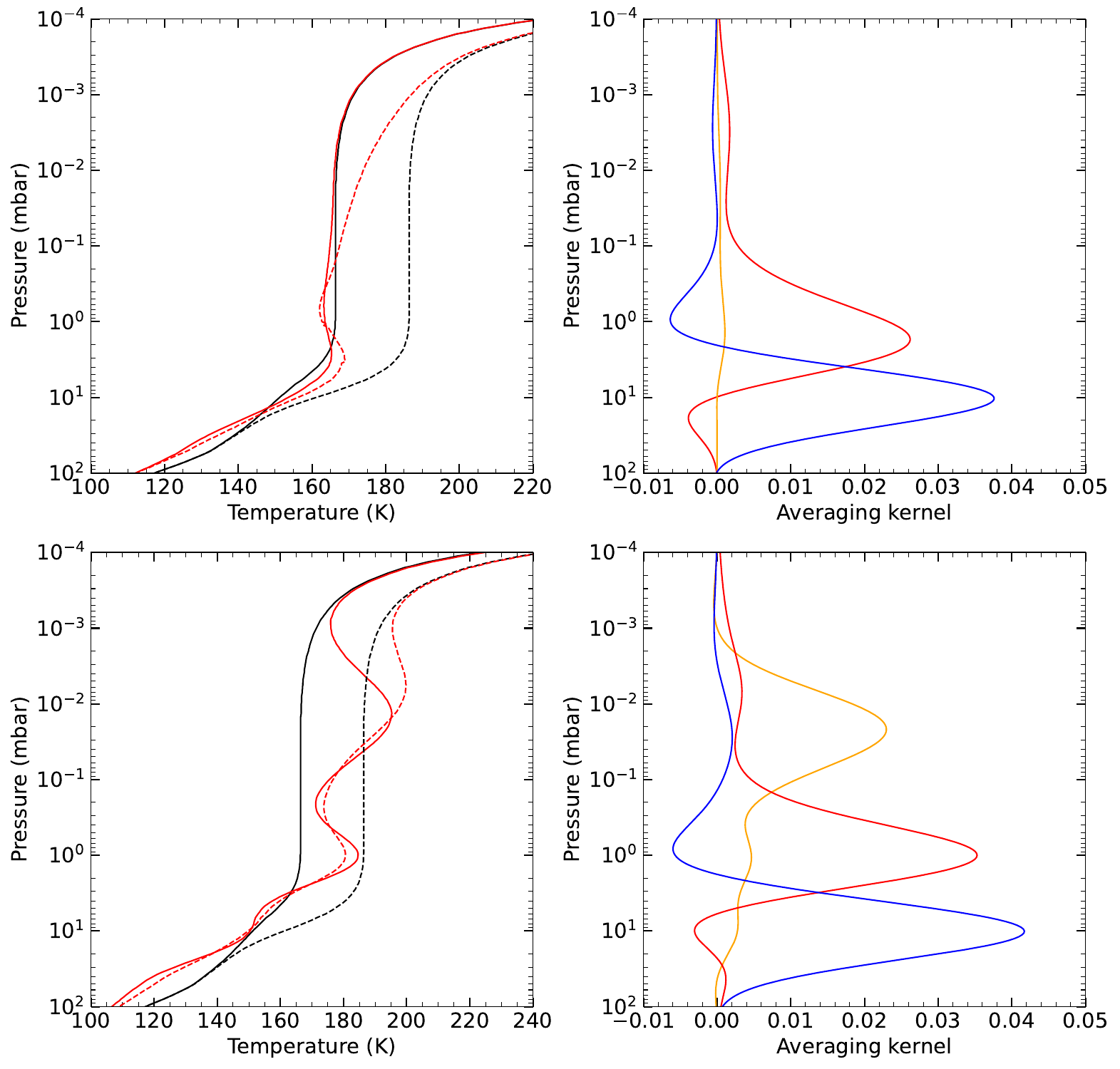}
    \caption{Left column: Comparison between the temperature profiles retrieved (red lines) outside (top) and inside (bottom) the auroral oval, with the \textit{a priori} profiles used (black lines). The continuous \textit{a priori} profile is the one used in \cite{Moses2005}, the dashed one is the same \textit{a priori} profile with a 20 K warmer region above 10 mbar to test the convergence of the fits. Right column: Averaging kernels for these thermal profiles at 20 (blue), 1 (red) and 0.01 mbar (orange).}
    \label{temp_profiles}
\end{figure}

The two left panels of Fig.~\ref{temp_profiles} present the inverted temperature profiles obtained using the CH$_4$ $\nu_4$ band (1240 --1330 cm$^{-1}$) for the two CH$_4$ spectra displayed in Fig.~\ref{all_spe}, sampling the thermal structure inside and outside the auroral oval. The inversion was carried out with a smoothing length of 0.75 scale height in the $S$ matrix, and using two different \textit{a priori} profiles (dashed lines in Fig.~\ref{temp_profiles}). The first is the temperature profile used in \cite{Moses2005} photochemical model, while the second \textit{a priori} profile deviates from the former, with a progressive increase starting at the 0.1-bar pressure level and being 20K warmer at the 1-mbar pressure level. 

Inspection of the averaging kernels displayed in the two right panels of Fig.~\ref{temp_profiles} confirms the sounded pressure levels. For the spectrum outside the auroral oval, significant averaging kernels are obtained up to the 0.5-mbar pressure level, while for the spectrum inside the auroral oval, the averaging kernels have significant values up to the 0.01 mbar level. This extra independent measurement compared to non-auroral regions is also illustrated by the number of degrees of freedom for each spectrum, 2.5 for the spectrum outside the auroral region, and 3.5 for the spectrum in the polar auroral region.

Using two different \textit{a priori} profiles enables us to confirm the vertical sensitivity offered by our spectra and to estimate the precision of our measurements, taking into account the uncertainties on the temperature profile beyond the sounded pressure levels. We note that for the spectrum taken outside the auroral oval, the two inverted temperature profiles coincide within the 30 -- 0.1~mbar pressure range. In contrast, for the spectrum sampling the interior of the oval, we can infer the temperature increase up to the 0.01 mbar pressure level, showing our capability to invert the temperature at higher altitudes within the auroral oval.

\begin{figure}[ht]
    \centering
    \includegraphics[width=\textwidth]{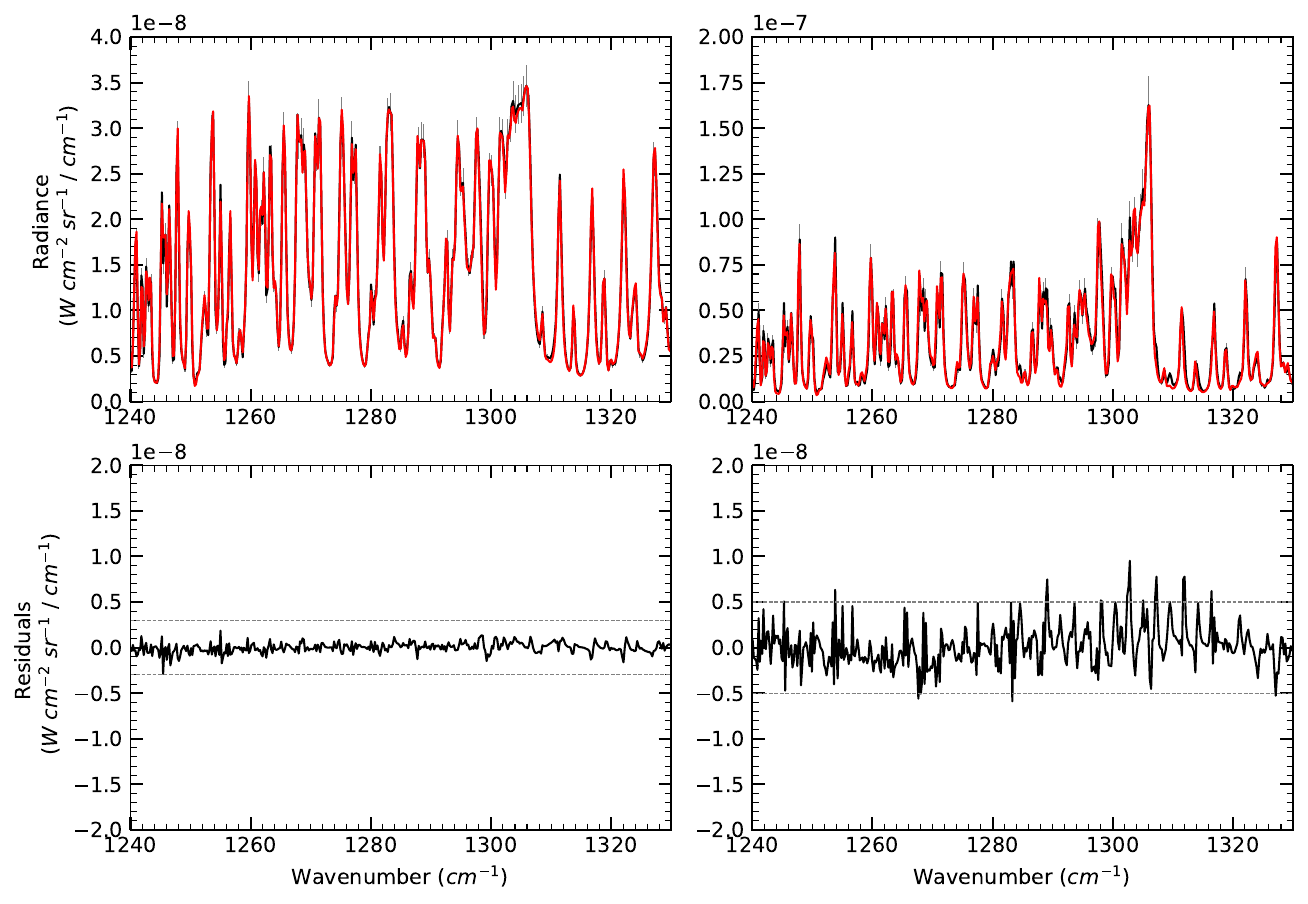}
    \caption{Top-Left: spectrum in the region of the CH$_4$ $\nu_4$ band of Jupiter outside the auroral oval at 61$^{\circ}$S, 63$^{\circ}$W in black, its best fit in red and the error bars in gray. Top-right: The same spectral band but for a region inside the auroral oval at 73$^{\circ}$S, 71$^{\circ}$W in black, its best fit in red and the error bars in gray. Bottom-Left: Residuals from the observation and the model for the spectra at 61$^{\circ}$S, 63$^{\circ}$W. Bottom-right: Residuals from the observed spectra and the model for the spectra at 73$^{\circ}$S, 71$^{\circ}$W. The dashed lines show the value of the 1$\sigma$ noise level. The vertical thermal profile from \cite{Moses2005} has been used as \textit{a priori}. Note the scales of the plots are different. }
    \label{spe_nu4}
\end{figure}

The goodness of the spectral fit is displayed in Fig.\ref{spe_nu4}. This figure shows that the difference in the shape of the Q-branch between the two spectra provides information on the temperature at high altitude. The residuals between the synthetic spectra and the observed spectra are larger for the scene inside the auroral oval. We think this may be due to molecular emissions not accounted for in our radiative transfer model. In particular, propane (C$_3$H$_8$) has several bands in this spectral region ($\nu_{12}$ and $\nu_{19}$) for which line lists are not included in HITRAN. Example of fits, information content and vertical sensitivity in the case of C$_2$H$_6$ and C$_2$H$_2$ retrievals will be shown in Section \ref{sec_res_hydroc}. 

\subsection{Error analysis}\label{sec_error}

The uncertainties affecting our retrieved temperature profiles are caused by two major sources. First, the instrumental noise level of MIRI-MRS. As explained in Sect.\ref{SecSNR}, two components contribute to MIRI's intrinsic noise: the Noise Equivalent Spectral Radiance (NESR), which is negligible for our dataset, and the calibration and reduction noise, which dominates the instrumental noise level. After adding the additional sources of error discussed in Section \ref{sec_observations}, the overall JWST NESR is translated into a precision better than 0.8 K on the retrieved temperature profiles.

The second major source of error is associated with uncertainties in the deep abundance of CH$_4$. In our analysis, we used a reference value of $1.9\times10^{-3}$, based on measurements from the Galileo probe \cite{Wong2004Sep}. We considered a range of variation for this abundance, from $1.5\times10^{-3}$ to $2.4\times 10^{-3}$, as Galileo Probe error bars span this VMR range. To assess the impact of this uncertainty, we repeated the full analysis for one hyperspectral cube by scaling the whole profile of the thirteen CH$_4$ models used to deep volume mixing ratios if $1.5\times10^{-3}$ and $2.4\times 10^{-3}$. We found that the uncertainty about the deep methane abundance has a negligible effect on the determination of the homopause height. On the other hand, we found significant variations in the inverted temperatures. Inside and outside the auroral oval, we found uncertainties of 1.5 K at both 1-mbar and 20-mbar pressure levels. Within the auroral region, the errors at the 0.01 mbar pressure level are larger, increasing up to 3 K.

For hydrocarbons, the main uncertainties in their vertical abundance profile result from the uncertainties in the temperature profile itself.

We also propagated the uncertainties on the telescope pointing into our retrievals. Since our observations are close to the limb of the planet, a small pointing error could strongly affect the calculated incidence and emission angles. To do so, we have performed several inversions by slightly varying the pointing specified in the metadata of the observations. We shifted the pointing, which is translated mainly in a change in latitude (of maximum $\pm$ 2$^{\circ}$ at 70$^{\circ}$S), and subsequently on the emission and incidence angles of each spaxel with respect to the unaltered scenario. Nevertheless, the  features observed after the inversion (see Section~\ref{sec_res_temp}) were the same in the two changed tests and in the unaltered spatial registration scenario in terms of spatial distribution. The largest change was located close to the limb, as the emission angle and latitude change rapidly due to the geometry of our observation. This test was performed to ensure that the features that will be shown in the following section are not related to incorrect spatial registration, but to robust atmospheric changes retrieved by our radiative transfer model.

\section{Results}\label{sec_results}
\subsection{Retrieval of the homopause height}\label{sec_homopause}

In this section, we present our indirect determination of the homopause pressure level through the CH$_4$ VMR profile, by analyzing the $\nu_4$ band of methane based on the procedure detailed in Sect. \ref{SecInformationContent}. Before presenting the results, we note some limitations that affect our determination of the homopause altitude from the CH$_4$ VMR profile.

\begin{figure}
    \centering
    \includegraphics[width=\textwidth]{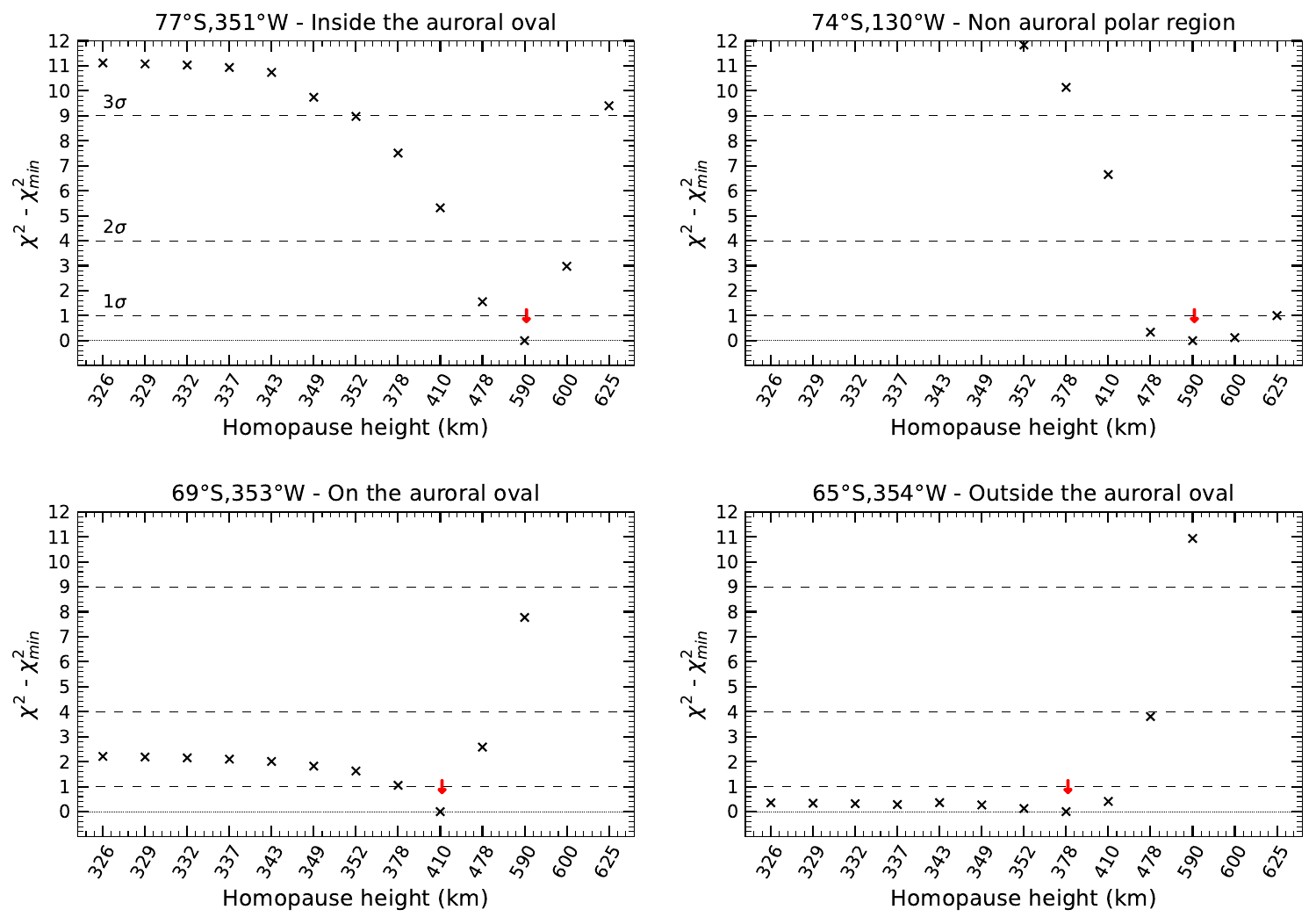}
    \caption{Variations in $\chi^2$ for the 13 homopause models obtained for different locations of the polar region. The red arrow shows the model which better fits the observation. The dashed lines show the levels of confidence of $1\sigma$, $2\sigma$ and $3\sigma$.}
    \label{chi2}
\end{figure}

\begin{figure}[t]
    \centering
    \includegraphics[width=\textwidth]{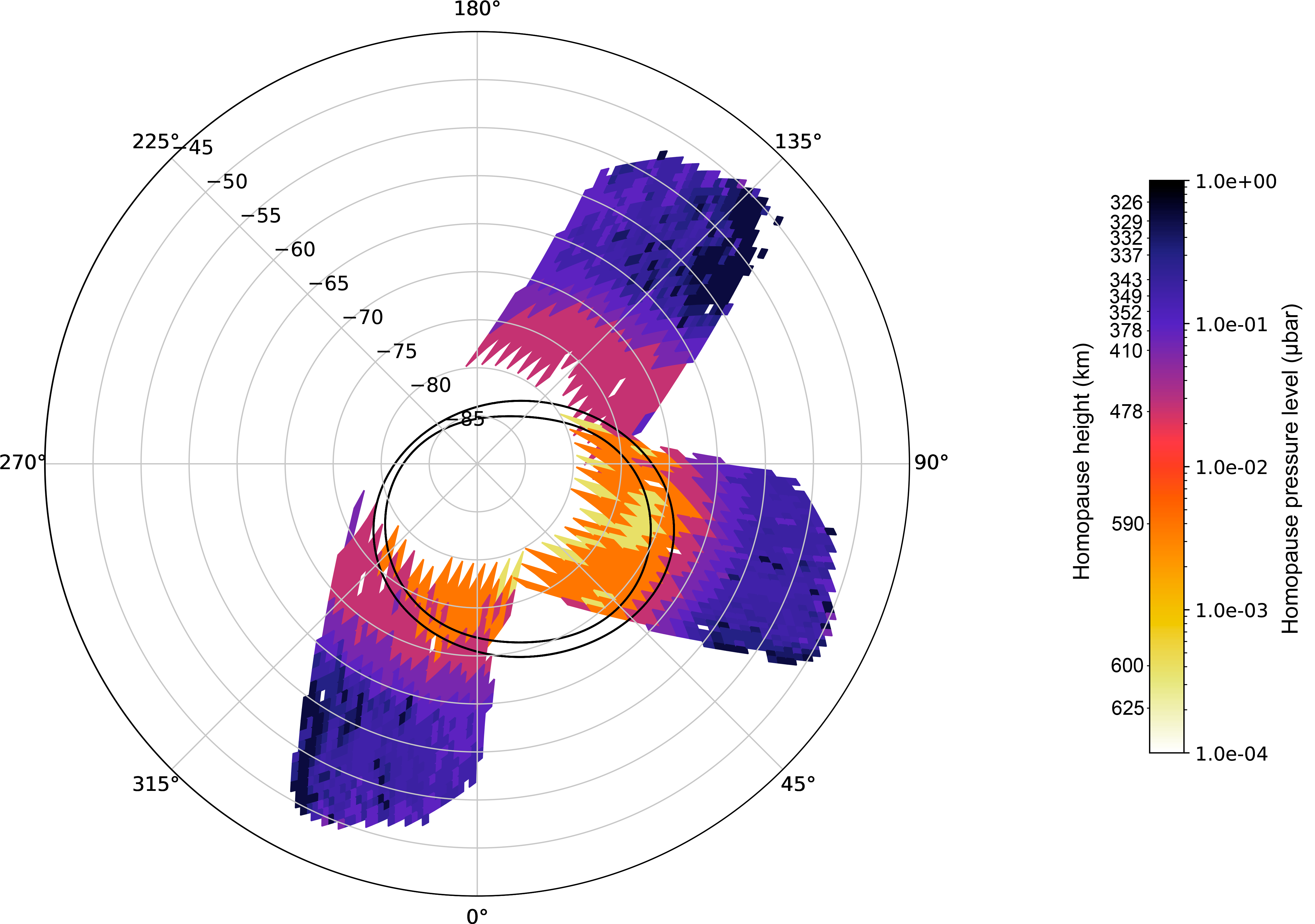}
    \caption{Polar projection of the homopause height retrieved for the three observations, using the $\nu_4$ band as spectral information. The black lines show the statistical position of the inner and outer borders of the auroral oval.}
    \label{homopause_nu4}
\end{figure}

First, Fig.~\ref{chi2} illustrates the $\chi^2$ values obtained in the retrieval of the temperature as a function of the 13 homopause pressure levels for four specific spectra. These individual spectra represent observations within or near the auroral oval, as well as those taken at quiescent latitudes equatorward of 65$^{\circ}$S. This figure shows that a well-defined $\chi^2$ minimum is achieved for spectra within or near the auroral oval, allowing a robust determination of the CH$_4$ profile and, in consequence, of the homopause height in this region. However, at latitudes equatorward of 65$^{\circ}$S, the $\chi^2$ lacks of a clear minimum, allowing us to establish only an upper limit of the homopause level, at a pressure of 0.3 $\mu$bar or greater for the majority of the spectra (corresponding to model $\#$8). This upper limit is presented on the homopause height map in Fig.~\ref{homopause_nu4} and was used for the temperature inversion in Sect.~\ref{sec_res_temp}. The reverse situation is encountered in a small specific area of the cube centered at 140$^{\circ}$W. Within a filament that extends meridionally from 130$^{\circ}$W to 155$^{\circ}$W, and centered at 74$^{\circ}$S, the $\chi^2$ as a function of the homopause height also lacks of a clear minimum, but this time it allows us to establish only a lower limit of the homopause level (see the upper right panel of Fig. \ref{chi2} for the spectra at 74$^{\circ}$S and 130$^{\circ}$W). The lower limit within this filament, $\sim$420 km ($\sim$60 nbar), actually corresponds to the determined homopause height in the surrounding area. Assuming this lower limit as the homopause height also yields a consistent temperature inversion with the surrounding area. In contrast, assuming a higher homopause level would produce a cold filament embedded within a warmer surrounding environment at the 1-mbar pressure level.

Furthermore, within the region spanning from 70$^{\circ}$S to 75$^{\circ}$S and 90$^{\circ}$W to 100$^{\circ}$W, two of our observations overlap and our analysis presents inconsistent homopause pressure levels. We tend to favor the higher homopause results derived from the cube centered at 70$^{\circ}$W due to its lower emission angles compared to those obtained from the cube centered at 135$^{\circ}$W. This preference is based on a more favorable nadir geometry, typically providing a more accurate estimation of the optical path length.

Figure~\ref{homopause_nu4} presents the homopause pressure levels retrieved for our three Jupiter regions, displayed in polar projection. The statistical position of the auroral oval is depicted by the black line for the day of our observations (December 24, 2022). The figure reveals a clear feature, showing that the homopause height rises southward at polar latitudes. The rise is the largest within the oval, where the homopause is found located at pressure levels as high as $\sim$ 0.4 -- 4 nbar ($625^{+2.5}_{-17.5}$ -- $590^{+17.5}_{-56}$ km), but it is also present at high pressure levels ($\sim$67 nbar or $\sim$410 km), outside the oval, e.g., southwards  70$^{\circ}$S. Inspection of the variation of the homopause pressure level along the 70$^{\circ}$S parallel clearly reveals this feature. Its lowest level of $\sim$ 91 nbar ($378^{+16}_{-13}$ km) is found at 150$^{\circ}$W, then it gradually increases to $\sim$ 0.4 nbar ($625^{+2.5}_{-17.5}$ km) at 70$^{\circ}$W, remaining around in the $\sim$ 0.4 -- 4 nbar range while inside the oval, and finally dropping to the $\sim$50~nbar ($410^{+34}_{-16}$ km) pressure level between 0$^{\circ}$ and 315$^{\circ}$W.

This behavior is also evident in the homopause altitude meridional gradient. The homopause is located at the highest altitudes at 70$^{\circ}$W, where the auroral oval reaches its lowest latitude. In comparison, at 135$^{\circ}$W and 300$^{\circ}$W, the homopause height drops more smoothly from high altitude within the oval to low altitude at quiescent latitudes. Equatorwards of 65$^{\circ}$S, the atmosphere appears to be relatively unaffected by auroral precipitation, as the homopause pressure level exhibits a relatively homogeneous altitude both zonally and meridionally. Variations of the homopause level may still exist at these latitudes, but our analysis can only constrain an upper limit of the homopause altitude in those regions.

Our results constitute the first spatially-resolved measurement of the homopause altitude in the SPR with enough information within the Southern Auroral Oval. Therefore, we can compare our retrievals only with studies that targeted the NPR. Our results in the SPR are qualitatively consistent with the study of \cite{Sinclairhomopause} who reported that the homopause in the Northern Auroral Oval lies at higher altitude than at mid-northern latitudes. These authors also reported that the contrast in homopause altitude between inside and outside the auroral oval decreased with increasing latitude. Thanks to the JWST angular resolution, such a trend is evident in our SPR map. Quantitatively,  \cite{Sinclairhomopause}  measured a homopause height located at $461^{+147}_{-39}$ km inside the Northern Auroral Oval, while our measurements yield a homopause located at $\sim$$590^{+17.5}_{-56}$ within the Southern Auroral Oval. 

\subsection{Temperature analysis}\label{sec_res_temp}

In this section, we present the results of the temperature structure analysis determined from our dataset. As mentioned in Sect.~\ref{sec_inversion}, the CH$_4$ $\nu_2$ band cannot be used to retrieve the thermal structure of the upper stratosphere due to its low sensitivity to high altitudes. Subsequently, it is not possible to retrieve the homopause height using this band. Hence, the results presented here were obtained by inverting the CH$_4$ $\nu_4$ band alone, using the corresponding CH$_4$ profile selected in \ref{sec_homopause}. 

\begin{figure}[ht!]
    \centering
    \includegraphics[width=\textwidth]{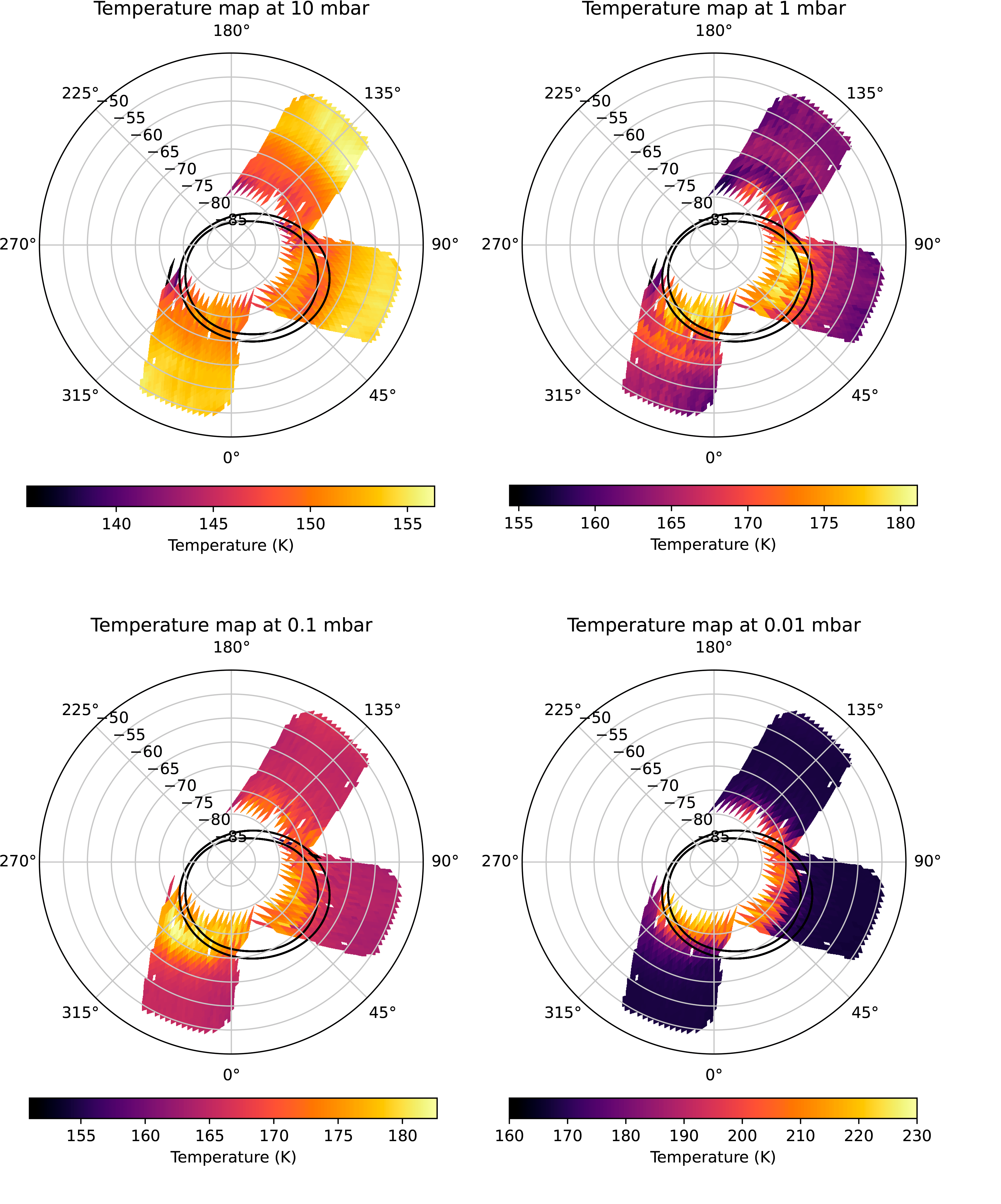}
    \caption{Polar projection of the temperature maps retrieved for the three observations at four different heights, using the $\nu_4$ band. The black lines show the statistical position of the inner and outer borders of the auroral oval.}
    \label{temp_maps}
\end{figure}

Figure \ref{temp_maps} presents the retrieved temperature for the three tiles of our mosaic at four different pressure levels, 10 mbar, 1 mbar, 0.1 mbar and 0.01 mbar using the $\nu_4$ band. For each spaxel, the displayed temperature correspond to the inverted profile using the homopause pressure level displayed in Fig.~\ref{homopause_nu4}. The four pressure levels were chosen to show the independent temperature measurements accessible within our sensitivity pressure range between 30 mbar and 0.01 mbar (Fig.~\ref{temp_profiles}).

At the 0.01-mbar pressure level, the temperature field displays a sharp polar warming southward of 70$^\circ$S, temperatures rising by an average of 37 $\pm$ 3 K from 175 K at 70$^\circ$S to 212 $\pm$ 3K at 80$^\circ$S. The warmest measured temperatures are located within the auroral oval at 78$^\circ$S and 350$^\circ$W, and the meridional temperature gradient also exists outside the auroral oval at 135$^\circ$W, albeit milder than within the oval. The gradient appears to decrease further eastward of 135$^\circ$W where the auroral oval retreats toward the pole, but unfortunately we lack a fourth tile around 225$^\circ$W to firmly assert this trend. We also note that the temperature field seems to be unaltered equatorward of 70$^\circ$S, the latitude corresponding to the equatormost extension of the auroral oval. 

At the 1-mbar pressure level, the temperature field morphology is similar to that at the 0.01-mbar level, but with a lesser contrast, of about 12 $\pm$ 2 K between 75$^\circ$S and 65$^\circ$S. If the polar warming is stronger within the auroral oval, as it is at 0.01-mbar, it also affects longitudes to the west of the auroral oval. However, we note that the return to an undisturbed temperature field is found to occur at more equatorward regions than at 0.01-mbar level.

At 0.1-mbar, the situation is slightly different. The temperature field displays a mild (10 K) polar warming similar in its longitudinal profile at 1 mbar. But it also displays a strong warming outside the auroral oval, with the maximum measured temperature of 180 $\pm$ 2K located at 72$^\circ$S and 320$^\circ$W, around 7 K warmer than the mean temperature inside the auroral region.

Furthermore, the temperature field at the 10-mbar pressure level is drastically different. At this level, it presents a polar vortex of cold temperatures ($\sim$7 K colder) poleward of 65$^\circ$S. 



Our results present both similarities and differences with respect to previous investigations of the Jovian Polar Region. Consistent with the thermal fields measured both in the NPR and in the SPR by \cite{SinclairI,SinclairII,Sinclair2023Apr}, we find that:
\begin{itemize}
    \item The largest warming occurs at the 0.01-mbar pressure level within the auroral oval. Our maximum measured temperature of $218\pm3$ K is higher than the maximum temperature of $205\pm5$ K measured by \cite{SinclairI} and the $185\pm5$ K measured by \cite{SinclairII} in the NPR, and of $200\pm5$~K and $175\pm5$~K in the SPR. Our results are in better agreement with those of \cite{Sinclair2023Apr}, who measured a maximum of $210\pm5$K for the SPR. 
    \item The 1-mbar pressure level is the most aurorally affected level after the 0.01-mbar level, with the highest temperature still observed within the auroral oval. At this pressure level, we measure a maximum temperature of $178\pm2$ K, which is similar to the $175\pm3$ K retrieved by \cite{SinclairI} and $180\pm3$ K observed by \cite{Sinclair2023Apr}, but warmer than the $166\pm3$ K value measured by \cite{SinclairII}.
    \item Within the auroral oval, the vertical temperature profile exhibits a minimum at 0.1 mbar, showing a weak contrast between inside and outside the auroral region in our study, as well as in \cite{SinclairI,SinclairII}.
    \item The temperature field below the 2--3 mbar pressure level appears to be unaffected by auroral precipitation, showing a southward decrease, similar to that observed by \cite{Fletcher2016} and \cite{Bardet2022}. 
\end{itemize}
Regarding the differences from the previous studies, we note that
\begin{itemize}
    \item We measure a temperature enhancement at 1 mbar, at polar latitudes also outside the auroral oval, mostly to the west of the auroral oval. This warming was not observed by \cite{SinclairI} or \cite{SinclairII} in either the SPR or the NPR.
    \item At the 0.1-mbar pressure level (Fig.~\ref{temp_maps}, bottom left), we observe a strong warming to the east of the auroral region, never witnessed by \cite{SinclairI} and \cite{SinclairII}.
    \item At 10~mbar, the cold polar ring seen in our thermal map was not observed by \cite{SinclairI} and \cite{SinclairII} as they lacked spatial resolution to sample latitude southwards of 70$^{\circ}$S at pressure levels larger than 1~mbar.
    \item We also note that our inferred thermal structure differs significantly from the one retrieved during a solar wind compression event by \cite{Sinclair2023Apr} for the SPR. In their study, the temperature increase within the auroral region was similar in the two hotspots at 1 and 0.01 mbar ($\sim 21 \pm 5$ K). This is in contrast to our observation, where the auroral hotspot presents the highest temperature at 0.01-mbar. Furthermore, \cite{Sinclair2023Apr} observed that compression also affected the structure of the temperature down to 10-mbar, while in our case the hotspot is only present at altitudes above the 5-mbar level.
\end{itemize}

\subsection{Hydrocarbons retrieval}\label{sec_res_hydroc}

To retrieve the ethane and acetylene volume mixing ratios, we have adopted the temperature structure and the homopause pressure based upon the CH$_4$ $\nu_4$ lines, as presented in the previous sections. For acetylene, we have analyzed the $\nu_5$ band centered at 730 cm$^{-1}$ covered in channels 3A and 3B, but restricted to the 685 -- 720 cm$^{-1}$ spectral range (channel 3B). We chose this specific wavenumber range for several reasons. First, as mentioned in Section \ref{sec_observations} the $\nu_5$ Q-branch radiance is saturated around 730~cm$^{-1}$. Second, we favored this short wavenumber range because it features the hot band $\nu_5 + \nu_4 - \nu_4$ that probes higher pressure levels than the $\nu_5$ band alone. We have also discarded wavenumbers close of 700 and 750~cm$^{-1}$, where the spectral signature of stratospheric aerosols was reported by \cite{Guerletaerosol} in Saturn's polar stratosphere.

To retrieve the abundance of ethane, we have used the $\nu_8$ band in the spectral range from 1510 to 1535 cm$^{-1}$, located in channel 1B next to the $\nu_2$ band of methane. The $\nu_8$ band extends further up to 1570 cm$^{-1}$, but we have excluded this range to avoid possible interferences between the C$_2$H$_6$ and the CH$_4$ $\nu_2$ lines.

The spectral features used to infer C$_2$H$_2$ and C$_2$H$_6$ abundances are located in different MRS channels (channels 1 and 3) than those used to retrieve the temperature and the homopause height (channel 2). Since each MIRI-MRS channel has a unique FOV, slice width, and pixel size, we needed to remap our inferred temperature structure and homopause height to align with the angular coverage and sampling of channels 1 and 3. For channel 3, which extends to higher northern latitudes than channel 2 due to its larger FOV, we assumed a constant temperature field and homopause height north of the limit of channel 2. In the inversion process of each spaxel, we adopted as \textit{a priori} C$_2$H$_2$ and C$_2$H$_6$ profiles the ones obtained using the \cite{Moses2017} photochemical model for our determined CH$_4$ homopause height, as the location of the homopause also affects to the vertical profiles of other chemical species such as C$_2$H$_2$ and C$_2$H$_6$.

\begin{figure}[t]
    \centering
    \includegraphics[width=\textwidth]{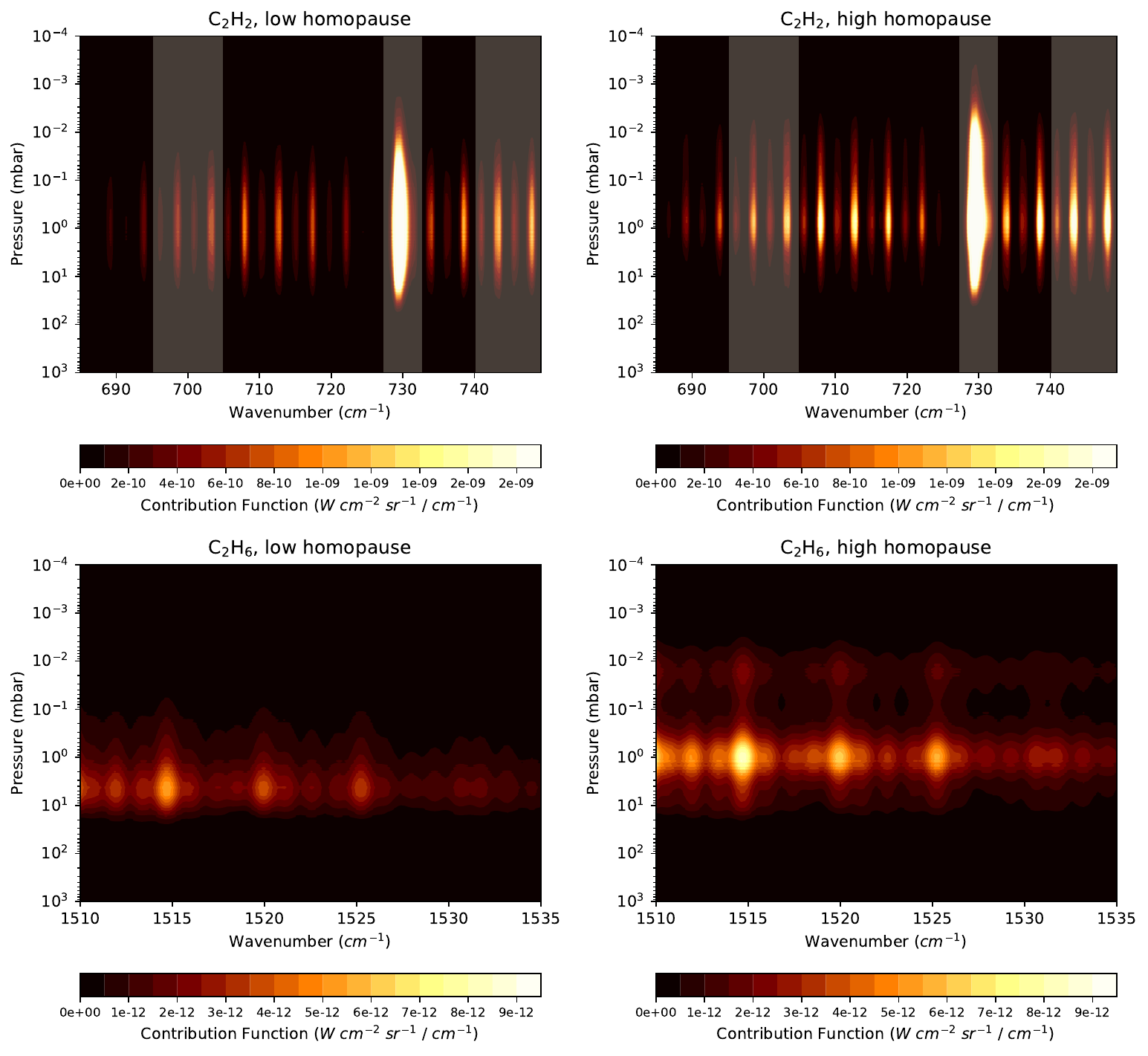}
    \caption{Contribution functions for C$_2$H$_2$ $\nu_5$ band (first row) and the C$_2$H$_6$ $\nu_8$ band (second row) for two different homopause locations. The spectral regions marked with the grey boxes in the C$_2$H$_2$ are skipped due to the saturation of the 730 cm$^{-1}$ feature, and the interference of the hazes (see Section \ref{sec_discussion}).}
    \label{hydroc_contfunc}
\end{figure}

Figure~\ref{hydroc_contfunc} shows the contribution functions in the respective spectral ranges for the two hydrocarbons, both inside and outside the auroral oval, i.e. for a high homopause and a low homopause altitude. The C$_2$H$_2$ contribution functions show that the acetylene bands analyzed provide information between 10 and 0.01~mbar, both inside and outside the auroral oval. In the case of C$_2$H$_6$, the information yielded by the $\nu_8$ band is concentrated between 10 and 1 mbar outside the auroral oval, while within the auroral oval, the sensitivity peak shifts slightly upwards, probing between the 10 and 0.1-mbar pressure levels. To account for the limited vertical sensitivity compared to the temperature sounding, we have adopted a different covariance matrix S (see Section~\ref{sec_inversion}) for the hydrocarbon inversion. While the smoothing factor was set to 0.75 scale heights for the temperature inversion, we have fixed it to 3 scale heights for the hydrocarbon inversion.

\begin{figure}
    \centering
    \includegraphics[width=\textwidth]{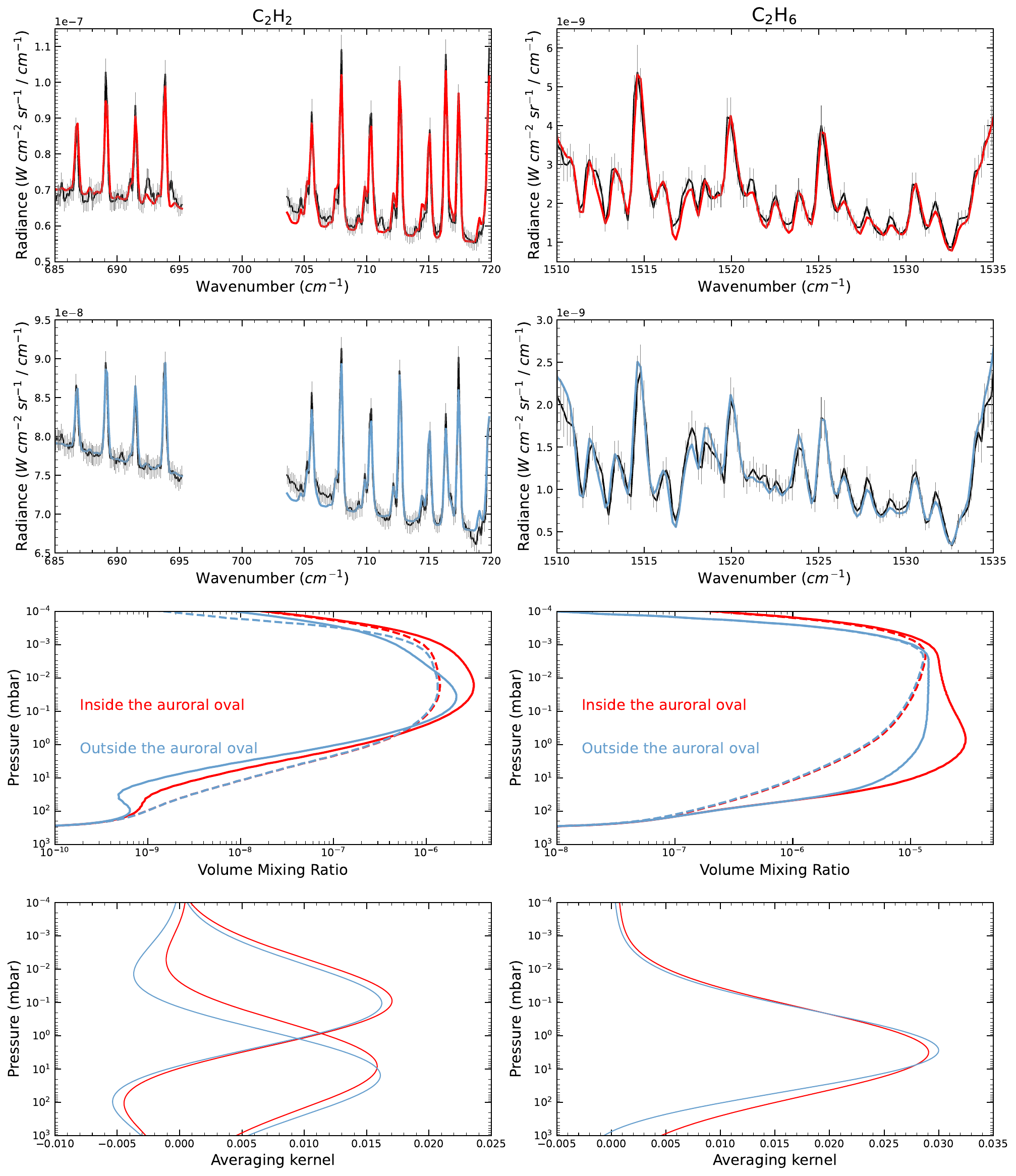}
    \caption{Top-left: Observed spectra of C$_2$H$_2$ (black) and the best fit (red) for a mean spectrum centered at 75$^{\circ}$S inside the auroral oval. Second row-left: Observed spectra of C$_2$H$_2$ (black) and the best fit (blue) for a mean spectrum located at 60$^{\circ}$S outside the auroral oval. Third row-left: Retrieved C$_2$H$_2$ vertical profile for the spectra inside the auroral oval (continuous-red) and outside the auroral oval (continuous-blue), with the corresponding \textit{a priori} profiles (dashed lines). Bottom-left: Averaging kernel for the two spectra of C$_2$H$_2$ at the two pressure levels to which the spectra is sensitive (7 and 0.1 mbar).
    Top-right: Observed spectra of C$_2$H$_6$ (black) and the best fit (red) for a mean spectrum centered at 75$^{\circ}$S inside the auroral oval. Second row-right: Observed spectra of C$_2$H$_6$ (black) and the best fit (blue) for a mean spectrum located at 60$^{\circ}$S outside the auroral oval. Third row-right: Retrieved C$_2$H$_6$ vertical profile for the spectra inside the auroral oval (continuous-red) and outside the auroral oval (continuous-blue), with the corresponding \textit{a priori} profile (dashed lines). Bottom-right: Averaging kernel for the two spectra of C$_2$H$_6$ for the pressure level to which the spectra is sensitive (3 mbar). For the observed spectra, the error bars are shown in gray.}
    \label{hydroc_spe_inv}
\end{figure}

Figure~\ref{hydroc_spe_inv} shows the comparison between the observed zonal-mean spectrum and the best-fit synthetic spectrum for the two example spectra shown in Fig.~\ref{all_spe}, representing both inside the auroral oval (top row) and one outside the auroral oval (second row). The third row displays the \textit{a priori} and inverted profiles for both inside and outside the auroral oval, while the bottom row presents the averaging kernels for the two regions. The left column shows results for C$_2$H$_2$, and the right column for C$_2$H$_6$. For C$_2$H$_2$, the averaging kernels show $\sim$2 degrees of freedom peaking at 7 and 0.1 mbar for all the spectra analyzed. On the other hand, for C$_2$H$_6$ we are able to retrieve its abundance peaking near 3~mbar for all the spectra, with $\sim$1 degree of freedom. 

\begin{figure}[t!]
    \centering
    \includegraphics[width=\textwidth]{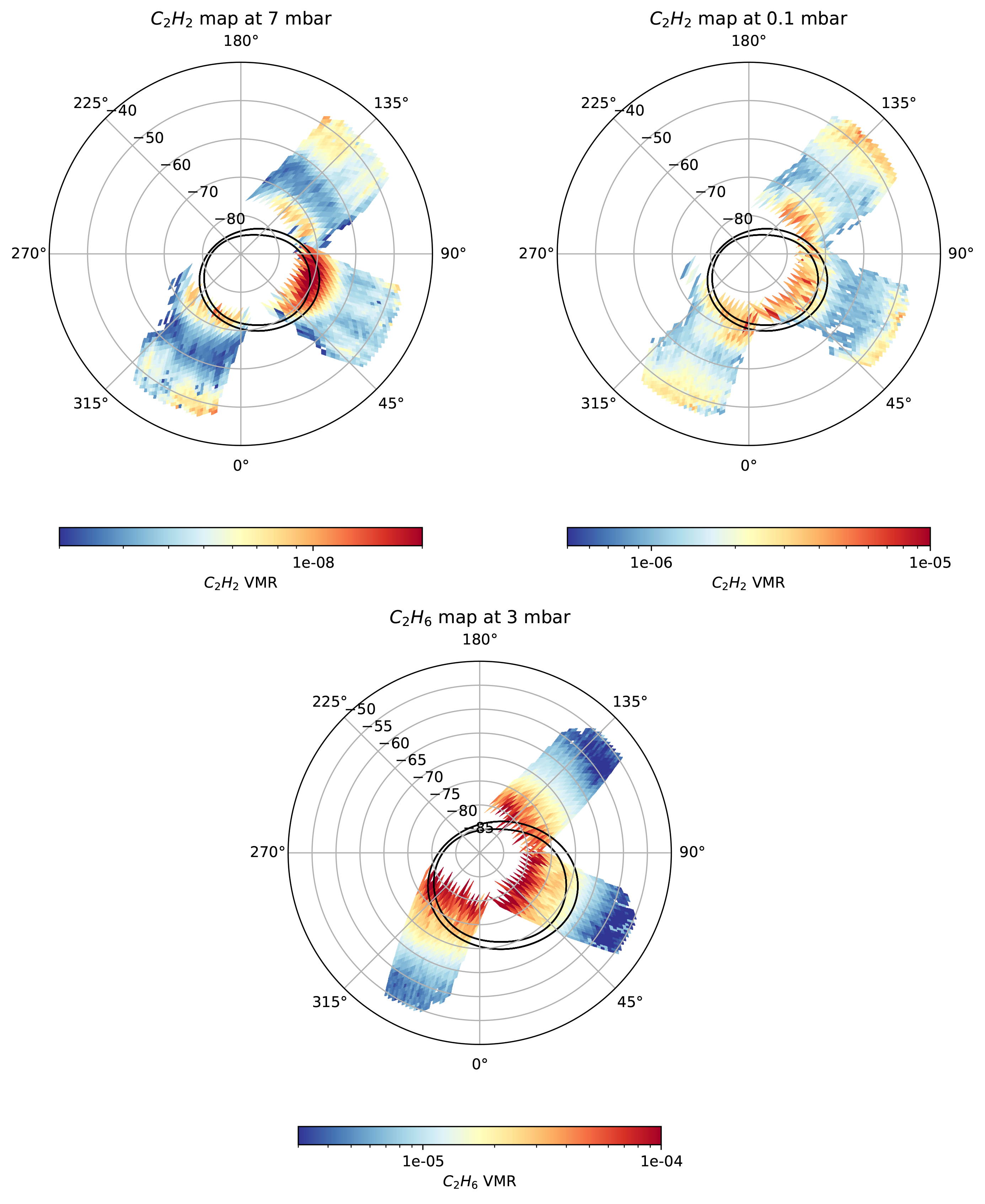}
    \caption{Top row: Polar projection of the C$_2$H$_2$ abundance at two pressure levels (7 and 0.1 mbar) for latitudes between 45$^{\circ}$S to 90$^{\circ}$S.
    Bottom: Polar projection of the C$_2$H$_6$ abundance at 3 mbar for latitudes between 55$^{\circ}$S to 90$^{\circ}$S.
    The black lines show the statistical position of the inner and outer borders of the auroral oval.}
    \label{hydroc_maps}
\end{figure}

Fig.~\ref{hydroc_maps} displays the inverted abundance of ethane and acetylene for the three mosaic tiles in polar projection at 5 and 0.1 mbar for C$_2$H$_2$, and at 3~mbar for C$_2$H$_6$.

\subsubsection{C$_2$H$_2$}\label{sec_res_c2h2}

The two spectra displayed in Fig.~\ref{hydroc_spe_inv} (left column) show that the difference in absolute radiance at the core of the emission lines is approximately 20 $\%$ higher inside compared to that outside the auroral oval (60$^{\circ}$S). 
Such a difference in line-to-continuum contrast cannot be explained just by the warmer auroral temperatures, but must also be the signature of an increase of the C$_2$H$_2$ abundance. This is demonstrated by the difference in the two inverted profiles displayed in the third row, left column panel of Fig.~\ref{hydroc_spe_inv}, where the profile inside the aurora (thick red line) is always larger than the profile outside the auroral region (thick blue line). 

As already pointed out by \cite{SinclairII,SinclairIII}, we are unable to achieve a fully satisfactory fit of the acetylene lines. The emission in the core of the lines is underestimated by our model, especially within the auroral oval. \cite{SinclairII} suggests that this may be caused by non-LTE effects occurring in the upper atmosphere, 
as the cores are sensitive to higher altitudes. At these altitudes, non-LTE phenomena may occur, since the thermal collisional timescale can become longer than the spontaneous radiative lifetime of the $\nu_5$ band. Using the methods described in \cite{Manuel-book,Sanchez-Lopez2022Jun}, we have estimated the deviation of the vibrational temperature from the kinetic temperature for the C$_2$H$_2$ $\nu_5$ band. We found that the excitation of $\nu_5$ level by solar pumping, e.g. by absorption of solar radiation by the $\nu_5$+$\nu_9$ band near 3~$\mu$m, is negligible in comparison with the collisional excitation in the whole thermosphere. Further, we found that the vibrational temperature of the $\nu_5$ level starts to deviate (becoming smaller than the kinetic temperature) at pressure levels around 0.5 $\mu$bar and this depletion increases rapidly at higher pressure levels becoming about 100 K at 0.1 $\mu$bar. Hence, assuming non-LTE would lead to, if any, a decrease of the LTE radiance rather than an enhancement. We then think that the inability to fit the measured spectrum may come from the uncertainty (an underestimation) of the temperature profile in the pressure range 0.1--1~$\mu$bar. We note that even if the C$_2$H$_2$ $\nu_5$ level is underpopulated with respect to LTE in that region, its non-LTE population still depends significantly on the kinetic temperature. This explanation seems  plausible as auroral processes can drastically augment the thermospheric temperature and thus increase the radiance in the cores of the C$_2$H$_2$ lines. Another possible explanation of the differences between the synthetic and observed spectra was proposed by recent analysis of the MRS observations of Saturn. \cite{Fletcher2023Sepsaturn} have shown that these differences could also be explained by a poor characterization of the spectral resolution for the complete wavenumber range of the MIRI-MRS instrument \cite{Jones2023Aug}.

The abundance map displayed in Fig.~\ref{hydroc_maps} extends over our entire spatial coverage, spanning the differences between the two representative spectra highlighted above. Globally, the meridional trend of acetylene shows a southward decrease in abundance at all pressure levels from a local maximum at 50$^{\circ}$S of latitude (VMR of $(8\pm 1)\times 10^{-9}$ at 7 mbar and $(3\pm 0.5)\times10^{-6}$ at 0.1 mbar) to a local minimum around 60$^{\circ}$S (VMR of $(5\pm 0.35)\times 10^{-9}$ at 7 mbar and $(1\pm 0.15)\times10^{-6}$ at 0.1 mbar). Beyond this, there is a subsequent rise in abundance by up to a factor of 5 at the southernmost latitudes covered by our observations (with a VMR of $(2\pm 0.45)\times 10^{-8}$ at 7 mbar and $(5 \pm 0.5) \times 10^{-6}$ at 0.1 mbar). At both pressure levels, the polar maximum is evident within the auroral oval and extends to the west of the auroral oval, specifically around 135$^{\circ}$W longitude, gradually diminishing at the westernmost edge of our observations. In particular, the most significant contrast between inside and outside the auroral oval is observed at the 7-mbar pressure level, where the abundance within the oval is twice as large as outside the oval.

\subsubsection{C$_2$H$_6$}\label{sec_res_c2h6}

Figure~\ref{hydroc_spe_inv} (right column) shows the comparison between the observed spectrum and the best-fitting synthetic spectrum for both inside the auroral region (top row) and outside the auroral region (second row). Similar to C$_2$H$_2$, there is an enhancement in C$_2$H$_6$ emission observed within the oval compared to outside. Taking the emission line at 1515 cm$^{-1}$ as a reference, and disregarding the continuum variations due to limb darkening, the increase amounts to approximately a factor of 2.



The retrieved ethane abundance map at 3 mbar is displayed in the lower panel of Fig.~\ref{hydroc_maps}. Unlike what we observed for C$_2$H$_2$, the C$_2$H$_6$ meridional trend is a monotonic increase in abundance as we approach the South Pole. Moreover, this increase is zonally uniform, and in particular there is no noticeable difference between regions inside and outside the oval at polar latitudes. Quantitatively, the VMR is found to increase by a factor of $\sim$7 from $(7\pm1.5)\times10^{-6}$ at $\sim$ 60$^{\circ}$S to $(4\pm 0.6)\times10^{-5}$ at $\sim$ 75$^{\circ}$S.

\section{Discussion of the results}\label{sec_discussion}
\subsection{Homopause}\label{sec_disc_homopause}

In our analysis and radiative transfer code, we assumed LTE emission at all pressure levels. However, it is known that LTE breaks down at higher altitudes, where density decreases. A study by \cite{Sanchez-Lopez2022Jun} on CH$_4$ $\nu_3$ non-LTE emission reveals that the vibrational temperature of various CH$_4$ vibrational levels begins to deviate from the kinetic temperature at pressure levels around 0.1 $\mu$bar. This deviation becomes substantial at pressures lower than 50~nbar. As illustrated in Fig.~\ref{contfunc}, a significant portion of the emission occurs at levels beneath or near the pressure levels where the LTE assumption becomes invalid. Hence, we argue that our results are minimally affected by the LTE assumption due to the predominance of emission occurring beneath or close to pressure levels where LTE breaks down. We also note that a higher homopause level than that assumed by \cite{Sanchez-Lopez2022Jun} will rise the altitude at which LTE breaks up, since self-absorption tends to maintain LTE conditions.

While our LTE assumption may introduce uncertainties in the absolute values of our inferred homopause level, it does not compromise the relative variations of the inferred homopause level. A larger CH$_4$ abundance at high altitude always results in higher opacity and more intense emission in the core of the CH$_4$ Q-branch. Some of our inferred homopause levels lie above the level where LTE breaks up. These homopause levels correspond to CH$_4$ abundances at pressure levels where the emission is in LTE, consistent with CH$_4$ vertical profiles for which the homopause level lies above the LTE break-up level.

At latitudes equatorward of 65$^{\circ}$S, i.e.\ away from the auroral oval, we could only establish an upper limit for the homopause height of $352^{+13}_{-3}$~km. These results are similar to those found by \cite{Sinclairhomopause}. For similar latitudes in the northern hemisphere, \cite{Sinclairhomopause} also determined an upper limit of the homopause height (see their Fig.~10), of $378^{+16}_{-13}$~km. The consistency between our upper limit in the southern hemisphere, with that determined by \cite{Sinclairhomopause} in the northern hemisphere suggests that, equatorward of 60$^{\circ}$N and 60$^{\circ}$S, the homopause height upper limit does not vary significantly with latitude across the planet's disk. Our upper limit is also consistent with the homopause altitude derived from stellar occultation observed by the Alice instrument onboard the New Horizons spacecraft, in the ultraviolet range. From this data set, \cite{Greathouse2010} deduced a homopause altitude of 310 and 340~km at the ingress (32$^{\circ}$N) and egress (18$^{\circ}$N) points, respectively. 

As we approach the South Pole, we observe an increase of the homopause height between latitudes 65$^{\circ}$S and 70$^{\circ}$S: southwards of 70$^{\circ}$S, the homopause altitude is higher than $410^{+34}_{-16}$ km at all longitudes. This is also similar with the results of \cite{Sinclairhomopause}, who also measured a higher homopause in non-auroral north polar regions than at mid-latitudes, but located at lower altitudes (378 $^{+16}_{-13}$ km) than in our results. Moreover, as \cite{Sinclairhomopause} found for the NPR, we find that the homopause height in the SPR reaches its maximum within the auroral oval. The inferred homopause altitude in the South Auroral Oval is higher than in the Northern counterpart; we derived a value of 590 $^{+17.5}_{-56}$ km within the oval, with a maximum of 625 $^{+2.5}_{-17.5}$ km, both larger than the altitude of 478 $^{+56}_{-34}$ km derived by \cite{Sinclairhomopause} for the North Auroral Oval.

We note that our measurements of the homopause altitude within the auroral region for the North and South poles overlap within the error bars. A possible explanation for these distinct homopause altitudes may lie in the different shape and size of the two auroral ovals. The ratio between the North Auroral Oval and the South Auroral Oval areas is approximately 1.84. If the total precipitating energy integrated over the two regions is similar, this surface difference may result in a higher density of precipitating energy in the SPR compared to that in the NPR, leading to stronger atmospheric perturbations. However, the topology and amplitude of the magnetic field are asymmetrical between the north and south; hence the energy input by particle precipitation is not strictly identical for both polar regions. It also becomes difficult to measure the average precipitating energy, since it is a process strongly variable in time at different temporal scales. 

Our dataset hence demonstrates that the upward shift of the homopause altitude previously found for the Northern Auroral Oval also occurs in the Southern Auroral Oval, clearly linking this upward shift to auroral activity. However, our dataset does not allow us to disentangle the mechanisms postulated to cause this upward shift: enhancement in the eddy diffusion, or vertical advection \cite{Sinclairhomopause}. Within the auroral ovals, vertical advection speeds of up to 10 m s$^{-1}$ have been predicted in the 1--0.01~$\mu$bar pressure range  by 3D thermospheric circulation models \cite{Yates2020}. With such speeds, the homopause altitude could vary by hundreds of kilometers on timescales comparable to a Jovian day. In contrast, we note that the diffusion timescale $\tau = \frac{H^2}{K}$ for the typical CH$_4$ molecular coefficient of $2\times10^7$~cm s$^{-2}$ at 0.1~$\mu$bar \cite{Moses2005} is longer than a Jovian day: $\tau\sim10^6$~s for a scale height of 45~km. If the predicted vertical speed actually takes place in the upper Jovian atmosphere, vertical transport seems a more efficient mechanism than turbulence to raise the homopause level.

The strong temporal variation of auroral and precipitation processes may offer an alternative explanation for the discrepancies between the Southern Auroral Oval and the Northern Auroral Oval measurements obtained by \cite{Sinclairhomopause}. In response to this variability, the homopause altitude could dynamically adjust over time, and the respective measurements may represent instantaneous snapshots of the homopause level. 
Such a variability in the homopause level is also compatible with the advection and diffusion timescales, of the order of $3\times10^4$ and $10^6$~s, respectively. Depending on the dominant process, the homopause level could adjust rapidly, i.e.\ within a few Jovian days, in response to changes in energy input. It is worth noting that our observations did not coincide with a compression event in the Jovian magnetosphere (see supplementary material), phenomena known to intensify auroral precipitation \cite{Sinclair2023Apr}. However, energy enhancement can also arise from internal processes within the Jovian magnetosphere.

At polar latitudes, but outside the auroral oval, we observed an elevation in the homopause level compared to quiescent latitudes. This increase could be attributed to auroral precipitation occurring outside the main auroral oval, or to the horizontal diffusion of auroral energy and associated disturbances (such as enhanced hydrocarbons) from the main oval.

Bright patches of far-ultraviolet (FUV) emissions are observed very regularly equatorward of the main auroral emission, often appearing in clusters, and are linked to injections of hot plasma moving inward within the Jovian magnetosphere \cite{Nichols2023Oct}. Strong particle precipitations also occur at the moons' footprints equatorward of the main oval. The main auroral emission is most of the time brighter than the diffuse equatorward emission, reflecting higher energy input, but the latter covers a much larger area than the former. Moreover, the more intense electron precipitation do not coincide with the highest energy fluxes \cite{Gerard2014}. In the Juno UVS observations of Jupiter's auroral emissions, the diffuse emission can extend up to 70$^{\circ}$S, or even 65$^{\circ}$S \cite{Greathouse2021} in the SPR. These latitudes correspond nicely with the latitudes at which we observed the transition between the unperturbed homopause level that prevails at mid-latitudes, and the perturbed homopause level in the polar region. In this interpretation, the smaller elevation outside the main oval would reflect the lower-energy precipitation occurring there.


To assess whether the elevated homopause outside the main oval could be due to horizontal diffusion, we compare the timescale of lateral transport with the timescale of CH$_4$ profile vertical relaxation. We quantified the horizontal eddy diffusion coefficient ($K_{yy}$) consistent with the vertical diffusion coefficient ($K_{zz}$) for each model. We assumed the Eddy diffusion timescale to be $\tau_k = \frac{H^2}{K_{zz}}$, and that it would be similar for horizontal eddy diffusion. We estimated $K_{yy}$ for a characteristic length L between (distance from a region (75$^{\circ}$S, 90$^{\circ}$W) inside the auroral oval to a region outside the oval, but with a higher homopause (75$^{\circ}$S, 135$^{\circ}$W), of around 14000 km) using $K_{yy} = \frac{L^2}{\tau_k}$. We obtain values for $K_{yy}$ ranging between $3\times10^{12}$ and $1\times10^{14}$ cm$^2$ s$^{-1}$ for homopause locations between 26 and 0.5 nbar. These values are much smaller than the values obtained by \cite{Lellouch2006Oct} and \cite{Griffith2004Jul} by fitting the meridional spread of the SL9 impact debris: $\sim$3$\times10^{11}$ cm$^2$ s$^{-1}$ at 0.1 mbar. It remains uncertain how this horizontal eddy coefficient $K_{yy}$ scales with pressure. \cite{Hue2018Jun} proposed several vertical scalings, from constant with pressure to proportional to the vertical variations of $K_{zz}$. Assuming the $K_{yy}$ at 0.1 mbar derived from post-SL9 species, combined with a vertical scaling following the one from $K_{zz}$, it would lead to a $K_{yy}$ coefficient of $\sim$2$\times10^{13}$ cm$^2$ s$^{-1}$ at 0.1 $\mu$bar, reading the $K_{yy}$ value at 0.1 $\mu$bar on Fig. 7 of \cite{Hue2018Jun} ("Lellouch 1" curve). This value would be sufficient to explain the horizontal spreading of the CH$_4$. However, \cite{Hue2018Jun} could not find which vertical scaling fits better the meridional trend of hydrocarbons. Furthermore, the presence of strong stratospheric jets at $\sim$0.1-mbar observed by \cite{Cavalie2021} may strongly hinder meridional mixing in the polar regions compared to the open stratosphere over the mid-latitudes probed by the dispersion of debris from SL9 impact. Therefore, it is difficult to draw any conclusions with regard to the actual physical process leading to the rise of the homopause outside the auroral region.

This elevated homopause may also help explain an apparent contradiction in observations of Jupiter’s 'swirl' aurora.  In the X-ray, the highest energy precipitation occurs on the dusk flank of the polar cap, with almost no emission within the ‘swirl’ region on the dawn side of the polar cap \cite{Dunn2020Jun}.  Similarly, regions with the most energetic precipitation are typically relatively weak in H$_3$$^+$ emission, as these particles penetrate well below the homopause; notably the 'swirl' region is relatively brightly observed in H$_3$$^+$ emission compared with UV emission, suggesting weaker precipitation flux \cite{Stallard2016Apr}. However, in the UV, the swirl region has very high color ratios, showing strong hydrocarbon absorption, which are typically aligned with highly energetic and deeply penetrating precipitation ($>400$ KeV; \cite{Gerard2014}). Here, we have shown that the homopause increases strongly in altitude in this polar region. This potentially allows much lower energy particle precipitation to penetrate into the elevated hydrocarbon layer, without necessarily driving significant ion production. Maps of C$_2$H$_2$ reflectance also appear to show the strongest enhancement in production in the dusk side, away from the 'swirl' region \cite{Giles2023Feb}, suggesting the 'swirl' region may not be a significant source of deeply penetrating high energy precipitation, and may instead be dominated by relatively low energy precipitation that is spectrally affected by the inflated atmosphere in this region.

Furthermore, a study of the CH$_4$ fluorescence at 3030 cm$^{-1}$ in the auroral region with JUNO-JIRAM data indicates that, although the homopause appears to be elevated in the auroral region, the increase in radiance may also be explained by higher temperatures in the nbar region \cite{Castagnoli2022Jul}. This correlation between localized homopause at high pressure levels with high temperatures in the upper atmosphere would indicate that in the auroral oval, the temperature could also be higher above 1 $\mu bar$.

\subsection{Thermal structure}\label{sec_disc_temp}

In the Southern Auroral Oval hotspot located at 0.01 mbar (see Fig.~\ref{temp_maps}), we retrieve atmospheric temperatures approximately 15 K $\pm$ 4 K warmer than those retrieved in the Northern Auroral Oval by \cite{SinclairII}. This increased auroral warming in the south is consistent with our hypothesis that the auroral energy density is larger in the SPR than in the NPR, proposed to explain the discrepancies between the homopause level in both regions. In addition, our observations were taken during a low solar wind activity period on Jupiter (see supplementary materials), meaning that the persistence of this warmer Southern Auroral Oval hotspot could be permanent, due to the smaller size of the Southern Auroral Oval compared to its northern counterpart. We also note that the JWST PSF at 7.7~$\mu$m is smaller than that of the IRTF, used by \cite{SinclairII}, in good seeing conditions: 0.25" against 0.5" at best. The dilution of the hotspot in the IRTF PSF could result in an underestimation of the actual temperature. At the 1-mbar pressure level, our temperature values are similar to those measured by \cite{Sinclair2023Apr} in the Southern Auroral Oval, with a difference of $12\pm$ 2K with respect to non-auroral regions. Moreover, the auroral warming we measure in the Southern Auroral Oval is quite similar to that inferred in the Northern Auroral Oval by \cite{SinclairI} and \cite{SinclairII}.

In between these two auroral warmings, we find a milder thermal contrast at the 0.1-mbar pressure between inside and outside the auroral regions. This is in line with findings from previous studies \cite{SinclairII}. We observe this milder contrast in the inverted profiles using the two a priori profiles (the \textit{hot} and \textit{cold} profiles) but we caution the reader that our spectra lack sensitivity at 0.1~mbar. This limitation is depicted in Fig.~\ref{temp_profiles}, where the averaging kernels exhibit weaker sensitivity at the 0.1-mbar level compared to the 0.01 and 1-mbar. Hence, we cannot confidently establish whether the auroral hotspots are actually separated by a colder layer or if our inverted temperature profile is relaxing to its $\textit{a priori}$ state.

Consistent with the interpretation of \cite{SinclairI}, we suggest that the pronounced warming at 0.01 mbar is the result of the auroral processes themselves. About the 1-mbar warming, \cite{SinclairI} consider it unlikely that it could be attributed to deep auroral precipitation or deep Joule heating, or alternatively to the conduction of auroral-induced warming from upper layers. They rather favored two other possibilities: i) warming by net radiative forcing of auroral-produced aerosols, or ii) adiabatic heating resulting from auroral-driven downwelling. We tend to prefer the latter, dynamical explanation. Indeed, \cite{Guerlet2020time}, using a radiative-equilibrium model of Jupiter's atmosphere, showed that the net radiative heating induced by polar aerosols was located in the lower stratosphere (10--30 mbar) and more zonally extended than the observations of a warming mostly restricted to regions inside the ovals. This situation is also observed from the analysis of near-infrared observations by \cite{Zhang2013Sep} that found that the large stratospheric aerosols reside in the 10--20 mbar pressure range at latitudes higher than 60$^{\circ}$S. \cite{Guerlet2020time} also noted that the current uncertainties on the aerosol shape, structure, and spectrometric parameters make a broad range of temperature profiles possible, from hardly any warming to a very strong, 40-K, warming. More recently, \cite{Sinclair2023Dec} also ruled out this hypothesis, as they found that the temperature at 1 mbar did not change according to the seasonally-changing
solar insolation due to Jupiter’s $\sim$3$^{\circ}$ axial tilt. Furthermore, images in several spectral ranges, from the UV to the near-infrared \cite{Barrado-Izagirre2008Mar}, reveal a haze layer more zonally and latitudinally extended than the auroral ovals themselves (see Fig. 1, filters F164N, 212N and 360M from \cite{Hueso2023Oct}). If the warming at 1 mbar were caused by aerosol forcing, it would be more zonally and latitudinally extended than observed. Hence, we rather support the dynamical origin of the 1-mbar warming. However, transient aerosol enhancements at 1--20 mbar levels inside the southern auroral oval may provide a mechanism for enhanced aerosol-related heating inside the oval. \cite{Tsubota2023Sep} observed a near-UV darkening inside the southern auroral oval, consistent with enhanced aerosols, 42 days prior to the JWST observations. The aerosol enhancement may have persisted over a significant fraction of the 42-day interval, given the 80--120 day lifetime of these polar UV-dark ovals from Cassini and Hubble Space Telescope data (\cite{Porco2003Mar}, \cite{Barbara2024Mar}, \cite{Tsubota2023Sep})

Recently, using ALMA, \cite{Cavalie2021} detected strong stratospheric jets at 0.1 mbar associated with the auroral oval, especially in the SPR where the geometry of observation was the most favorable. A strong jet was indeed predicted by \cite{Yates2020} using a General Circulation Model coupling the magnetosphere, ionosphere and thermosphere. In this model, the jet is associated with a downwelling and adiabatic warming needed to maintain the fast wind in hydrostatic equilibrium. Observations of an ionospheric/thermospheric jet with speeds an order of magnitude faster than the stratospheric jet seen with ALMA suggest that ion drag transfers momentum to the neutral atmosphere \cite{Wang2023Dec}. Rotation rates of UV-dark ovals in the deeper stratosphere (2--20 mbar) are likewise an order of magnitude weaker than the auroral stratospheric jets (\cite{Tsubota2023Sep}, \cite{Barbara2024Mar}). The vertical wind shear may indicate that the momentum transfer extends deep into the stratosphere within the auroral ovals, potentially modifying temperature profiles via enhanced eddy mixing.

A mechanism that could possibly extend the hotspot in the vertical direction from 0.01-mbar down to 1-mbar is atmospheric wave breaking. In this mechanism, the hotspot created at the higher atmospheric levels by auroral energy precipitation modifies the atmospheric stability so that gravity or Kelvin waves cannot propagate upward through the hotspot. Hence, they would deposit their heat and angular momentum just below the hotspot, where vertical static stability is changed the most compared to the quiescent atmosphere, and contributes to propagate it downwards, and drive the observed strong auroral jets. Such a mechanism is suggested to explain Saturn's equatorial oscillation \cite{Bardet2022} or stratospheric beacon \cite{Fletcher2012,Fouchettemp}.

The auroral hotspot is no longer present at the 10-mbar pressure level. In our dynamical interpretation, it means that a process damps the downward propagation of the hotspot. We propose that the auroral hotspot is damped by radiative cooling due to aerosols that accumulate at this pressure level, as suggested by \cite{Zhang2015Dec}. Indeed, the equatorward frontier of the cold polar vortex centered at 65$^{\circ}$S coincides with the northernmost extension of the aerosol polar cap (see Fig.~\ref{temp_maps}, and also \cite{Zhang2015Dec}). \cite{Guerlet2020time} have shown that aerosols can sharply affect the radiative equilibrium temperature in the polar regions, although large uncertainties on their radiative properties still exist and affect the predicted temperature profile. \cite{Guerlet2020time} also showed that aerosols can reduce the radiative time constant in the polar regions to as a short as $\sim100$~terrestrial days. Such a short time constant might explain how the stratosphere could react to compression events produced by strong heating at deep pressure levels and relax rapidly to the pre-compression state. This would match the observations of \cite{Sinclair2023Apr}, who found an increase of auroral temperatures as deep as the 10-mbar pressure level during a magnetic compression event. In contrast, our observations were performed during a minimum of solar wind compression (see supplementary materials) according to the model from \cite{Tao2005Nov}, which favors the hotspot being limited to higher altitudes.

\begin{figure}[t]
    \centering
    \includegraphics[width=\textwidth]{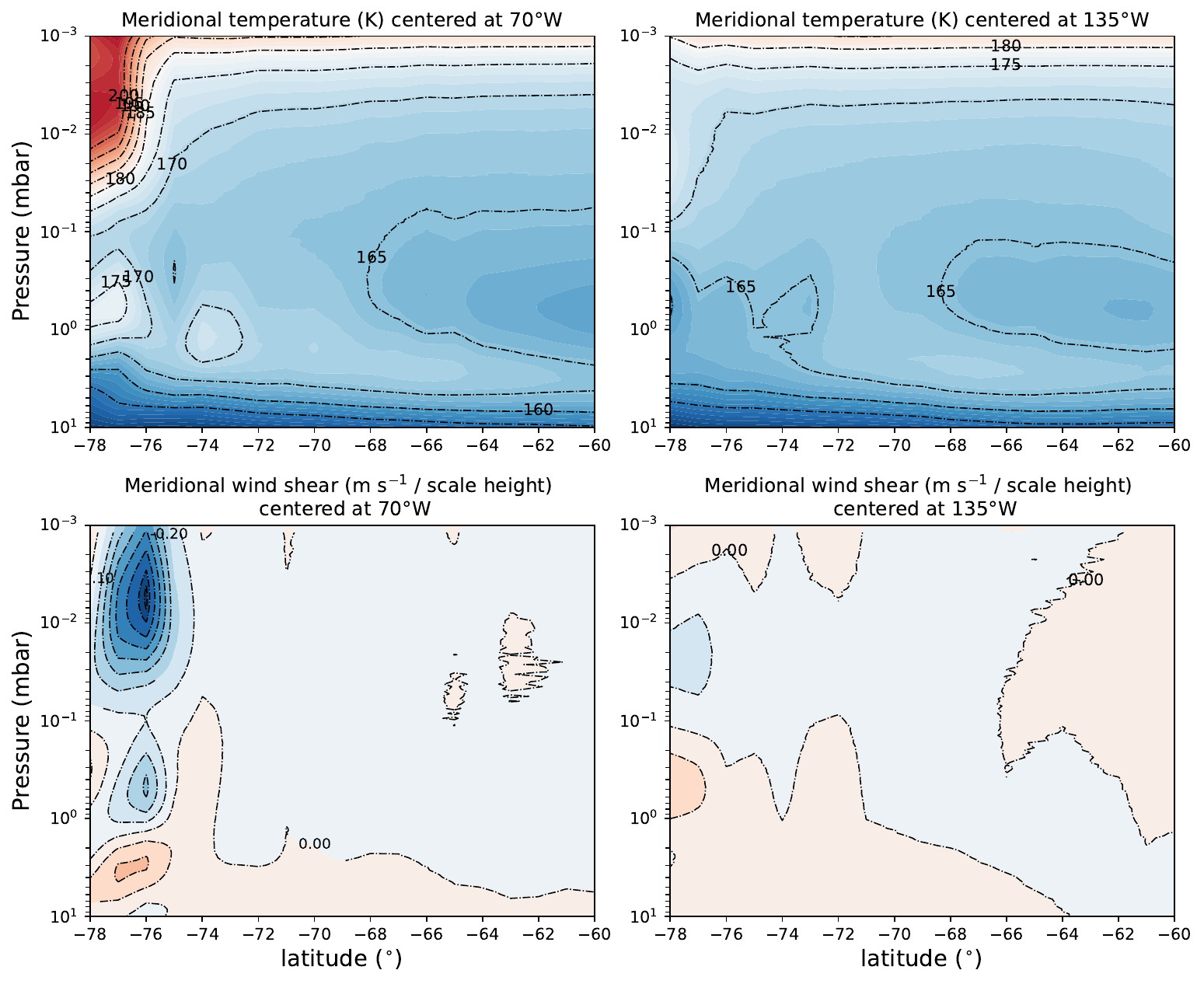}
    \caption{Top: Meridional temperature maps centered at 70$^{\circ}$W (left) and 135$^{\circ}$W (right). The auroral oval is only present in the left panel. The dashed lines are equally separated each 5K. Bottom: The wind shear for the top panels. The dashed lines are equally separated each 0.1 m s$^{-1}$ / scale height.}
    \label{wind_shear}
\end{figure}

\cite{Cavalie2021} did not find evidence for the auroral jet between 1 and 3 mbar. To better constrain the vertical structure of this jet, we have calculated the vertical shear of the zonal wind from the latitudinal thermal gradient of each tile in our observation. We have calculated it following the expression

\begin{equation}
    \frac{\partial u}{\partial z} = - \frac{g}{f T} \frac{\partial T}{\partial y} ,
\end{equation}

\noindent where u is the zonal wind, z the height, g is gravity, f is the Coriolis parameter, and T is temperature. Fig.~\ref{wind_shear} shows the meridional temperature maps for two tiles (centered at 70$^{\circ}$W and 135$^{\circ}$W, respectively), with and without the auroral oval inside the FOV. The bottom row shows the calculation of the wind shear for the upper meridional temperature maps. We observe that the values of the wind shear at these latitudes are close to zero, as previously reported by \cite{Fletcher2016}. Yet, we note that in the presence of the auroral oval, higher values of wind shear are obtained, at pressure levels similar to those where \cite{Cavalie2021} found the auroral jet. These wind shear values indicate a low variation of $\sim$8~m s$^{-1}$ per scale height, which would indicate that the jet located at 0.1 mbar with a zonal speed of -340 m s$^{-1}$ should extend from at least 10~mbar up to 0.01 mbar. Hence, this jet could encapsulate the two hotspots and serve as a mechanism that isolates the warmer auroral region from its surroundings. This increase in jet speed with altitude supports the fact that the jet observed by \cite{Cavalie2021} could be part of the electrojet \cite{Rego1999May} observed in the sub- $\mu bar$ regions by \cite{Chaufray2011Feb} with speeds of 1 -- 3 km s$^{-1}$. However, below 3 mbar, the wind shear map shows a decrease in zonal wind velocity. This seems to confirm the idea that this jet weakens at deeper levels, but the measured meridional gradients and inferred vertical shear do not seem compatible with a null wind at the 10-mbar pressure level. However, the fact that \cite{Cavalie2021} did not find evidence of this jet at 1 mbar, may indicate that this jet has a cutoff at lower pressures than 10 mbar, and that our analysis is limited by the low vertical resolution of the inverted temperature profiles. 

Our analysis suggests that some regions located outside the auroral ovals also experience some warming (especially at 1 and 0.1 mbar) but in a less intense way than within the oval. We note the reader, though, that the measurements of the homopause and temperature are slightly correlated and that it is difficult to accurately assess the magnitude of the warming in regions affected by a homopause rising. However, as already presented in Sect.~\ref{sec_disc_homopause}, auroral precipitation is also present outside the auroral oval \cite{Greathouse2021}, and may contribute to atmospheric warming in regions neighboring the auroral ovals.

This warming outside the oval affects our calculation of the wind shear. The larger meridional gradient between the temperature inside and outside the oval observed at 0.01 mbar contributes to a larger wind shear at this pressure level. At the 1-mbar level, however, the thermal contrast is not as pronounced and leads to milder wind shear that results in a slow downward decrease of the observed jets.

\subsection{Hydrocarbon abundances}\label{sec_hydroc_abund}

The meridional distribution of C$_2$H$_2$ at both 0.1 and 7 mbar (see supplementary materials) shows first a decrease in abundance starting from 55$^{\circ}$S towards the South Pole, with a localized minimum at 65$^{\circ}$S. From this latitude, C$_2$H$_2$ increases up to 3 times higher at 75$^{\circ}$S than at 55$^{\circ}$S in regions where the auroral oval is present. This trend is similar to that observed in \cite{Nixon2010Nov} from Cassini-CIRS observations, although with better spatial coverage of the polar latitudes. Due to the proximity of the Southern Auroral Oval to the polar axis, it is quantitatively difficult to compare our results with those of \cite{SinclairI, SinclairII}, as the Southern Auroral Oval was not clearly visible in their analysis. Compared to the results of \cite{Sinclair2023Apr} for the SPR, we find that our abundances are approximately a factor of 2 times smaller at 7 mbar, but 5 times higher at 0.1 mbar. This is quite surprising since the observations in \cite{Sinclair2023Apr} were performed during a strong solar wind event; hence, we should expect our abundances to be globally lower than theirs. The abundances within the Southern Auroral Oval are about 1.5 -- 2 times higher than those measured by \cite{SinclairI, SinclairII} for the Northern Auroral Oval, as well as the higher temperatures, could be related to possible higher energetic electron concentration in the Southern Auroral Oval compared to that in the Northern Auroral Oval. Nevertheless, in both polar regions the C$_2$H$_2$ enhancement suggests significant particle precipitation below the homopause, aligned with measurements of the spectral color ratio of auroral UV emission, which shows both the main auroral emission and especially the 'swirl' auroral region precipitation deep into the hydrocarbon layer (e.g. \cite{Gerard2014,Gerard2018Sep}).

The increase of C$_2$H$_2$ due to ion-neutral chemistry requires the dissociation of C$_2$H$_5$$^{+}$ from particle precipitation to generate C$_2$H$_2$, as proposed by \cite{Giles2021Aug}, based on previous work on Titan chemistry \cite{DeLaHaye2008a,Loison2015Feb,Dobrijevic2016Apr}:

\begin{equation}
    CH^{+}_3 + CH_4^{-} \rightarrow C_2H_5^{+} + H_2
\end{equation}
\begin{equation}
    C_2H^{+}_5 + e^- \rightarrow C_2H_2 + H_2 + H.
\label{c2h2_chem}
\end{equation}

Comparing the VMR inside the auroral oval with the minimum value located at $\sim$60$^{\circ}$S for the three tiles, we found that the contrast is about a factor of 2 higher at 0.1 mbar when compared with the contrast at 7 mbar, and that the highest contrast is found in the tile centered at 70$^{\circ}$W for both the 0.1-mbar and 7-mbar pressure levels. We suggest that the gradient in abundance from auroral latitudes down to 65$^{\circ}$S is due to transport from an auroral production site to a destruction site driven by neutral photochemical processes linked to insolation \cite{Hue2018Jun}. The fact that the gradient is weaker at 7 mbar than at 0.1 mbar suggests that the meridional transport by C$_2$H$_2$ diffusion is more efficient at deeper pressure levels. This is in agreement with the wind vertical decay of the auroral jet we inferred from the analysis of the thermal wind shear in Section \ref{sec_disc_temp}, and is in agreement with the lack of jet at 1--3 mbar reported by \cite{Cavalie2021}.



\cite{Giles2023Feb} also found that this minimum is located between 60 and 65$^{\circ}$S using Juno UVS observations. Their study of the SPR also showed an enhancement of C$_2$H$_2$ outside the oval (see their Fig 5, panels c and e), and longitudinally mixed at latitudes poleward 70$^{\circ}$S, with the peak values inside the auroral oval. We find that the abundance of C$_2$H$_2$ is also enhanced in the equatorward vicinity of the auroral oval, with its peak value inside the oval. This mixing outside the oval is not so evident in the NPR \cite{Sinclair2023Apr}, and appears to be due to local properties of the SPR, such as its smaller size (thereby higher particle precipitation density) or its proximity to the rotation axis, which could contribute to a more efficient zonal transport. Both factors would generate an increase in C$_2$H$_2$ production and would facilitate its mixing with non-auroral regions. However, a more conclusive explanation of the transport mechanism regarding C$_2$H$_2$ is still needed and is beyond the scope of this work.

The current knowledge of the abundance and distribution of C$_2$H$_6$ in the polar regions of Jupiter is more complex than for C$_2$H$_2$ or C$_2$H$_4$. While in \cite{SinclairI} the C$_2$H$_6$ was shown to be depleted at 5 and 1 mbar, in \cite{SinclairII} this abundance was shown to be enriched at 5 mbar. However, in both studies, this difference was not strong enough to consistently constrain the behavior of C$_2$H$_6$ in the NPR. In \cite{Sinclair2023Apr}, despite the better quality of the observations, it was not possible to distinguish any spatial distribution related to auroral activity.

In our C$_2$H$_6$ retrieval at 3~mbar, we find a poleward enhancement, but not localized within the auroral oval. Our results show an enhancement of the C$_2$H$_6$ abundance by a factor of 8 at 75$^{\circ}$S compared to the C$_2$H$_6$ at 55$^{\circ}$S. The retrieved meridional trend is in contrast to the observation of \cite{SinclairI, SinclairII} , who invoked a recycling of C$_2$H$_2$ into C$_2$H$_6$ outside the auroral oval, where neutral photochemistry prevails, resulting in an apparent depletion of C$_2$H$_6$ inside the auroral oval. We note that the enhancement of C$_2$H$_6$ as we approach the polar regions is not localized inside the auroral oval and has previously been observed at lower latitudes \cite{Nixon2010Nov}. This could imply that ion-neutral chemistry would not affect C$_2$H$_6$ in the same way it does C$_2$H$_2$, and that C$_2$H$_6$ is mainly controlled by photochemistry rather than by ion chemistry.

We cannot state that ion-neutral chemistry does not affect the production of C$_2$H$_6$, as the longer lifetime of this molecule compared to C$_2$H$_2$ could be a factor to explain why at 3 mbar this molecule seems to be longitudinally well mixed.

We found that the auroral production of C$_2$H$_6$ may be due to the following reactions described by \cite{DeLaHaye2008a}: 

\begin{equation}
    C_2H^{+}_5 + C_3H_8 \rightarrow C_2H_6 + C_3H_7^{+}
\label{c2h6_chemx}
\end{equation}
\begin{equation}
H^{+}_3 + C_4H_{10} \rightarrow C_2H^{+}_5 + C_2H_6 + H_2
\label{c2h6_chem}
\end{equation}

According to \cite{SinclairII}, these C$_2$H$_6$ production reactions (\ref{c2h6_chemx},\ref{c2h6_chem}) are approximately two orders of magnitude less efficient than C$_2$H$_2$ production reaction (\ref{c2h2_chem}) in the auroral region. Although the increase of C$_2$H$_6$ at 75$^{\circ}$ compared to a region located at 50$^{\circ}$ is less than that of C$_2$H$_2$ by a factor of 2 -- 3, it is far from the ratio yielded by the efficiency of the chemical reactions. However, we note that this chemical efficiency is based on photochemical models. The abundance of C$_3$H$_8$ in the polar regions of Saturn is increased by a factor of 2 with respect to mid-latitudes \cite{Guerlet2009Sep}. If charged particle precipitation on Jupiter is more efficient than on Saturn in terms of hydrocarbon production, we could expect a higher increase in C$_3$H$_8$, which could be the source of C$_2$H$_6$ in the auroral region. Unfortunately, we cannot accurately measure the abundance of C$_3$H$_8$ given the current state of the JWST data reduction pipeline and the not yet definitive characterization of the MIRI-MRS spectral resolution, which makes the retrieval of the abundance of C$_3$H$_8$, which shares an emission line with C$_2$H$_2$ at 730 cm$^{-1}$, complex. 

 From a dynamical point of view, the C$_2$H$_6$ abundance map may suggest that, as observed for aerosols in \cite{Zhang2013Sep}, this hydrocarbon would have been able to escape the auroral jet at 3 mbar and have been efficiently mixed throughout the polar region. However, the fact that we find  C$_2$H$_2$ enhanced inside the auroral oval at lower pressure levels (7-mbar) seems to indicate that the dynamical and chemical processes related to the hydrocarbons in the auroral regions are more complex than explained from previous work. 


\section{Conclusions}\label{sec_conclusion}

JWST Mid Infrared Instrument - Medium Resolution Spectrometer observations of Jupiter's South Polar Region have allowed us to perform a detailed analysis of its polar stratosphere. A total of three tiles were obtained on 24 December 2022. This observation covered latitudes from 50$^{\circ}$S to 84$^{\circ}$S. The three exposures were centered at 340$^{\circ}$W, 70$^{\circ}$W and 140$^{\circ}$W, and in the 340$^{\circ}$W and 70$^{\circ}$W the auroral oval was visible. 

We performed temperature retrieval tests using two different methane bands ($\nu_2$ and $\nu_4$). Since Jupiter's upper atmosphere is affected by auroral precipitation phenomena, the $\nu_2$ band, which probes lower pressure levels ($\sim$ 1 mbar), is not effective for analyzing the thermal structure of Jupiter's stratosphere. We have used the $\nu_4$ band of CH$_4$, using different atmospheric models for different homopause altitudes, obtaining an elevated homopause of at least 200 km within the Southern Auroral Oval (590$^{+17.5}_{-56}$~km) compared to atmospheric regions not affected by particle precipitation, where an upper limit of 349~km could only be retrieved for the altitude of the homopause. This shows that the southern auroral region experiences the same upward shift of the homopause as the northern auroral region \cite{Sinclairhomopause}. We tend to favor vertical advection as the most efficient mechanism to transport CH$_4$ to higher altitudes, this process being triggered by the energetic injection of auroral particles into the atmosphere. The homopause also seems to be elevated at high latitudes outside the auroral oval, suggesting efficient zonal transport at high altitudes. These results do not take in consideration non-LTE effects happening at the pressure levels where the homopause is located. Hence, even though the retrieved homopause altitude may not be quantitatively correct, the data seems to indicate that it is located at higher altitudes within the auroral oval, even if we take into account non-LTE effects.

We have found two temperature peaks located at 1 and 0.01 mbar, very similar to those found in \cite{Sinclair2023Apr}. The 0.01 mbar peak is thought to be a lower altitude extension of the warmer thermosphere inside the auroral oval, and is hotter than those found in the North Polar Region \cite{SinclairI,SinclairII, Sinclair2023Apr}, possibly due to a higher density of deposited energy given the smaller size of the Southern Auroral Oval compared to its northern counterpart. Following \cite{SinclairII}, we favor adiabatic heating resulting from auroral-driven downwelling as the origin of the 1-mbar peak, since the high altitude auroral region seems to be confined by the auroral jet observed by \cite{Cavalie2021}. In addition, at 10 mbar, we observed a cold polar vortex with its equatormost region located at 65$^{\circ}$S. We propose that this may indicate the presence of stratospheric aerosols. 

Regarding the abundance of hydrocarbons, for C$_2$H$_2$ we found a decrease as we approached the pole, as predicted by photochemical models \cite{Hue2018Jun}. However, within the auroral oval, we found an increase of a factor of $\sim$3 compared to the abundance at 55$^{\circ}$S. C$_2$H$_6$ also shows an increase at 3 mbar when we approach the polar region. This is the first time that this behavior has been clearly observed in Jupiter's South Polar Region, as in previous studies the Southern Auroral Oval was not visible \cite{SinclairI, SinclairII} or the retrieved abundance did not show any characteristic trend \cite{Sinclair2023Apr}. This also demonstrates the complexity of the chemistry in this region: while C$_2$H$_2$ is enhanced at 7 mbar inside the auroral oval, C$_2$H$_6$ seems to be mixed throughout the polar region. Production of C$_2$H$_6$ in the auroral region could be triggered by the formation of C$_3$H$_8$ and other hydrocarbons, based on analogy with Titan's chemistry. Understanding the extent of the C$_2$H$_6$ produced from these species likely produced in Jupiter auroral region requires a new chemical model of Jupiter that account for the full range of ion-neutral chemistry, combined with constraints on the charged particles' precipitation. This task is beyond the scope of this paper, and should be addressed in future work.

The magnetosphere-atmosphere coupling at Jupiter is subject to strong temporal and dynamical variations, which makes understanding the effect of this coupling on the atmospheric chemistry complex. Complementary IRTF-TEXES and JWST observations of the North Polar Region could shed some light on the chemical processes occurring in the auroral regions, as well as potentially observe some asymmetries between the two hemispheres related to asymmetrical processes occurring in the magnetosphere of the planet. Future analysis of the aerosol budget in the auroral regions and their impact on the thermal retrievals will also allow us to better comprehend the dynamics and chemistry of the polar region of Jupiter.

\section*{acknowledgments}
This work is based on observations obtained with the NASA/ESA/CSA James Webb Space Telescope. The data were obtained from the Mikulski Archive for Space Telescopes at the Space Telescope Science Institute, which is operated by the Association of Universities for Research in Astronomy, Inc., under NASA contract NAS 5-03127 for JWST. These observations are associated with program $\#$1373 (Observations 2, 4 and 26), which is led by co-PIs Imke de Pater and Thierry Fouchet. Data from JWST programs 1246 and 1247 were used for wavelength calibration. PRO was supported by an Université Paris-Cité contract. TC acknowledges funding from CNES and from the Programme National de Planétologie. VH acknowledges support from the French government under the France 2030 investment plan, as part of the Initiative d’Excellence d’Aix-Marseille Université – A*MIDEX AMX-22-CPJ-04. MLP acknowledges financial support from the Agencia Estatal de Investigación, MCIN/AEI/ 10.13039/ 501100011033, through grants PID2019- 110689RB- I00 and CEX2021- 001131-S. IdP and MHW are in part supported by the Space Telescope Science Institute grant JWST-ERS-01373. JAS was supported by grant NNH17ZDA001N issued through the Solar System Observations Planetary Astronomy program., under a contract with the National Aeronautics and Space Administration (NASA) to the Jet Propulsion Laboratory, California Institute of Technology. JAS and GSO carried out some of this research at the Jet Propulsion Laboratory, California Institute of Technology, under a contract with the National Aeronautics and Space Administration (80NM0018D0004). LNF, OK and MTR were supported by a European Research Council Consolidator Grant (under the European Union's Horizon 2020 research and innovation programme, Grant 723890) at the University of Leicester. JH was supported by an STFC studentship; HM was supported by an STFC James Webb Fellowship (ST/W001527/1). RH and ASL  were supported by grant PID 2019-109467GB-I00 funded by MCIN/AEI/ 10.13039/501100011033/ and were also supported by Grupos Gobierno Vasco IT1742-22.

\section*{Open Research}

Level-3  calibrated  Jupiter  MIRI/MRS  data  from  the  standard  pipeline  are  available  directly  from  the  MAST  archive (https://mast.stsci.edu/portal/Mashup/Clients/Mast/Portal.html MISSION: JWST, PROPOSAL-ID: 1373). 

The radiative transfer and retrieval code used in this work and previous works \cite{Fouchettemp, Guerlet2009Sep} is available for download \cite{retrieval_git}.

The JWST calibration pipeline is available via \cite{pipeline_bushouse}, this work used version 1.11.3. The  data  products  produced in  this  study (temperature and abundance maps) are available from \cite{BibEntry2024Apr}.






\end{document}